# Measurement of activation cross-section of long-lived products in deuteron induced nuclear reactions on palladium in the 30-50 MeV energy range

F. Ditrói*[1], F. Tárkányi[1], S. Takács[1], A. Hermanne[2], A.V. Ignatyuk[3]

[1]Institute for Nuclear Research, Hungarian Academy of Sciences (ATOMKI), Debrecen. Hungary

[2]*Cyclotron Laboratory, Vrije Universiteit Brussel (VUB), Brussels, Belgium*

[3]*Institute of Physics and Power Engineering (IPPE), Obninsk, Russia*

Abstract

Excitation functions were measured in the 31 - 49.2 MeV energy range for the $^{nat}Pd(d,xn)^{111,110m,106m,105,104g,103}Ag$, $^{nat}Pd(d,x)$ $^{111m,109,101,100}Pd$, $^{nat}Pd(d,x)$, $^{105,102m,102g,101m,101g,100,99m,99g}Rh$ and $^{nat}Pd(d,x)^{103,97}Ru$ nuclear reactions by using the stacked foil irradiation technique. The experimental results are compared with our previous results and with the theoretical predictions calculated with the ALICE-D, EMPIRE-D and TALYS (TENDL libraries) codes.

* Corresponding author: ditroi@atomki.hu

# 1. Introduction

We have been continuing the investigation of activation cross-sections of proton, deuteron and alpha-particle induced reactions on palladium metal in connection with production of medically relevant radioisotopes (Ditrói et al., 2007; Hermanne et al., 2002, 2004a, b; Hermanne et al., 2004c; Hermanne et al., 2005; Tárkányi et al., 2009). Among others, the excitation curves and production capabilities for deuteron induced reactions on $^{nat}Pd$ for the medically relevant $^{104m,g}Ag$ (PET imaging), $^{111}Ag$ and $^{103}Pd$ (therapeutic applications) were investigated. The extension of cross-section database for proton induced reactions on palladium up to 80 MeV was recently published (Tárkányi et al., 2016). As for deuteron induced reactions no datasets above 40 MeV (up to 40 MeV only our earlier results) are available in the literature. So we extended the energy range of the experimental data up to 50 MeV in this work and included a comparison with theoretical calculations using different model codes.

## 2. Experiment and data evaluation

For the reaction cross-section determination an activation method based on stacked foil irradiation followed by γ-ray spectroscopy was used. The stack irradiated at Louvain la Neuve CYCLONE 110 cyclotron, consisted of a sequence of Rh (26 μm), Al (50 μm), Al (6 μm), In (5μm), Al (50 μm), Pd (8 μm), Al (50 μm), Nb (10 μm) and Al (50 μm) foils, repeated 9 times. Bombardment lasted for 3600 s with a 50 MeV deuteron beam of 32 nA.

The activity produced was measured non-destructively using HPGe γ-ray spectrometers. Five series of γ-spectra measurements were performed at different sample detector distances and for time intervals 8.3-10.0 h, 28.2-44.7 h, 173.8.4-243.0 h, 529.7-557.0 and 12667.3-12886.9 after end of bombardment (EOB), respectively. The evaluation of the γ-ray spectra was made by both an automatic and an interactive peak fitting code (Székely, 1985).

Effective beam energy and the energy scale were estimated initially by a stopping calculation (Andersen and Ziegler, 1977) based on the estimated incident energy and target thickness and finally corrected (Tárkányi et al., 1991) on the basis of the excitation functions of the $^{27}Al(d,x)^{22,24}Na$ monitor reactions (Tárkányi et al., 2001) simultaneously re-measured over the whole energy range**.** For estimation of the uncertainty of the median energy in the target samples and in the monitor foils, the cumulative errors influencing the calculated energy (incident proton energy, thickness of the foils, beam straggling) were taken into account.

The cross-sections were calculated from the well-known activation formula by using a custom developed computer algorithm that was used in many earlier evaluations (Ditrói et al., 2000; Ditrói et al., 2007). The results are given as production cross-sections, supposing the target is monoisotopic, containing contributions from reactions on the numerous stable Pd isotopes. The decay data were taken from the online database NuDat2 (NuDat, 2014) and the Q-values of the contributing reactions from the Q-value calculator (Pritychenko and Sonzogni, 2003), both are presented in Table 1. The beam intensity (the number of the incident particles per unit time) was obtained initially through measuring the charge collected in a short Faraday cup and corrected on the basis of the excitation functions of the monitor reactions compared to the latest version of the IAEA-TECDOC-1211 recommended database (Tárkányi et al., 2001). The uncertainty on each cross-section was estimated in the standard way (International-Bureau-of-Weights-and-Measures,

1993) by taking the square root of the sum in quadrature of all individual linear contributions, supposing equal sensitivities for the different parameters appearing in the formula: determination of the peak areas including statistical errors (1-20%), number of target nuclei including non-uniformity (3%), detector efficiency (5%) and incident particle intensity (7%). The total uncertainty on the cross-section values was evaluated to be approximately 10-20%.

.

Table 1 Decay data of the investigated reaction products (NuDat, 2014; Pritychenko and Sonzogni, 2003)

| **Nuclide<br>(Level)<br>Decay path** | **Half-life** | **$E_\gamma$(keV)** | **$I_\gamma$(%)** | **Contributing reaction** | **Q-value (keV) GS-GS** |
|---|---|---|---|---|---|
| **$^{111}$Ag**<br>$\beta^-$: 100 % | 7.45 d | 245.40<br>342.13 | 1.24<br>6.7 | $^{110}$Pd(d,n)<br>decay of $^{111m,g}$Pd | 4949.2<br>3501.7 |
| **$^{110m}$Ag**<br>117.595 keV<br>IT: 1.33 %<br>$\beta^-$: 98.67 % | 249.83 d | 657.76<br>677.62<br>706.68<br>763.94<br>884.68<br>937.49<br>1384.29 | 95.61<br>10.70<br>16.69<br>22.60<br>75.0<br>35.0<br>25.1 | $^{110}$Pd(d,2n) | -3880.7 |
| **$^{106m}$Ag**<br>89.667 keV<br>ε: 100 % | 8.28 d | 406.18<br>429.65<br>450.98<br>616.17<br>717.34<br>748.36<br>804.28<br>824.69<br>1045.83<br>1527.65 | 13.4<br>13.2<br>28.2<br>21.6<br>28.9<br>20.6<br>12.4<br>15.3<br>29.6<br>16.3 | $^{105}$Pd(d,n)<br>$^{106}$Pd(d,2n)<br>$^{108}$Pd(d,4n)<br>$^{110}$Pd(d,6n) | 3588.9<br>-5972.1<br>-21731.7<br>-36681.5 |
| **$^{105g}$Ag**<br>ε: 100 % | 41.29 d | 280.44<br>344.52<br>443.37<br>644.55 | 30.2<br>41.4<br>10.5<br>11.1 | $^{104}$Pd(d,n)<br>$^{105}$Pd(d,2n)<br>$^{106}$Pd(d,3n)<br>$^{108}$Pd(d,5n)<br>$^{110}$Pd(d,7n) | 2740.3<br>-4353.9<br>-13914.8<br>-29674.4<br>-44624.2 |
| **$^{104m}$Ag**<br>6.9022 keV<br>IT: 0.07 %<br>ε: 99.93 % | 33.5 min | 996.1<br>1238.8 | 0.50<br>3.9 | $^{104}$Pd(d,2n)<br>$^{105}$Pd(d,3n)<br>$^{106}$Pd(d,4n)<br>$^{108}$Pd(d,6n)<br>$^{110}$Pd(d,8n) | -7285.6<br>-14379.7<br>-23940.7<br>-39700.2<br>-54650.0 |
| **$^{104g}$Ag**<br>ε: 100 % | 69.2 min | 623.2<br>863.0<br>925.9<br>941.6 | 2.5<br>6.9<br>12.5<br>25.0 | $^{104}$Pd(d,2n)<br>$^{105}$Pd(d,3n)<br>$^{106}$Pd(d,4n)<br>$^{108}$Pd(d,6n)<br>$^{110}$Pd(d,8n) | -7285.6<br>-14379.7<br>-23940.7<br>-39700.2<br>-54650.0 |
| **$^{103}$Ag**<br>ε: 100 % | 65.7 min | 118.74<br>148.20<br>266.86<br>1273.83 | 31.2<br>28.3<br>13.3<br>9.4 | $^{102}$Pd(d,n)<br>$^{104}$Pd(d,3n)<br>$^{105}$Pd(d,4n)<br>$^{106}$Pd(d,5n)<br>$^{108}$Pd(d,7n)<br>$^{110}$Pd(d,9n) | 1933.7<br>-15672.9<br>-22767.0<br>-32328.0<br>-48087.6<br>-63037.38 |
| **$^{111m}$Pd**<br>172.21 keV<br>IT: 100 % | 5.5 h | 172.18 | 46 | $^{110}$Pd(d,p) | 3501.7 |
| **$^{109}$Pd**<br>$\beta^-$: 100 % | 13.7012 h | 88.04<br>311.4<br>647.3 | 3.6<br>0.032<br>0.0244 | $^{108}$Pd(d,p)<br>$^{110}$Pd(d,p2n)<br>decay of $^{109}$Rh | 3929.0<br>-11020.8<br>-12845.1 |
| **$^{101}$Pd**<br>ε: 100 % | 8.47 h | 269.67<br>296.29 | 6.43<br>19 | $^{102}$Pd(d,p2n)<br>$^{104}$Pd(d,p4n) | -12796.2<br>-30402.9 |
| | | 565.98<br>590.44 | 3.44<br>12.06 | $^{105}$Pd(d,p5n)<br>$^{106}$Pd(d,p6n)<br>$^{108}$Pd(d,p8n)<br>$^{101}$Ag decay | -37496.9<br>-47057.9<br>-62817.5<br>-17674.9 |
| **$^{100}$Pd**<br>ε: 100 % | 3.63 d | 74.78<br>84.00<br>126.15<br>158.87 | 48<br>52<br>7.8<br>1.66 | $^{102}$Pd(d,p3n)<br>$^{104}$Pd(d,p5n)<br>$^{105}$Pd(d,p6n)<br>$^{106}$Pd(d,p7n)<br>$^{108}$Pd(d,p9n)<br>$^{100}$Ag decay | -21071.0<br>-38677.6<br>-45771.7<br>-55332.7<br>-71092.2<br>-28942.6 |
| **$^{105}$Rh**<br>$\beta^-$: 100 % | 35.36 h | 306.1<br>318.9 | 5.1<br>19.1 | $^{105}$Pd(d,2p)<br>$^{106}$Pd(d,2pn)<br>$^{108}$Pd(d,2p3n)<br>$^{110}$Pd(d,2p5n) | -2009.4<br>-11570.4<br>-27330.0<br>-42279.8 |
| **$^{102m}$Rh**<br>140.7 keV<br>IT: 0.233 %<br>ε: 99.767 % | 3.742 y | 631.29<br>697.49<br>766.84<br>1046.59<br>1112.84 | 56.0<br>44.0<br>34.0<br>34.0<br>19.0 | $^{102}$Pd(d,2p)<br>$^{104}$Pd(d,2p2n)<br>$^{105}$Pd(d,2p3n)<br>$^{106}$Pd(d,2p4n)<br>$^{108}$Pd(d,2p6n)<br>$^{110}$Pd(d,2p8n) | -2592.8<br>-20199.4<br>-27293.5<br>-36854.5<br>-52614.1<br>-67563.9 |
| **$^{102g}$Rh**<br>$\beta^-$: 22 *5* %<br>ε: 62.3 5 %<br>β+: 15.7% | 207.3 d | 468.58<br>739.5<br>1158.10 | 2.9<br>0.53<br>0.58 | $^{102}$Pd(d,2p)<br>$^{104}$Pd(d,2p2n)<br>$^{105}$Pd(d,2p3n)<br>$^{106}$Pd(d,2p4n)<br>$^{108}$Pd(d,2p6n)<br>$^{110}$Pd(d,2p8n) | -2592.8<br>-20199.4<br>-27293.5<br>-36854.5<br>-52614.1<br>-67563.9 |
| **$^{101m}$Rh**<br>157.41 keV<br>IT: 7.20 %<br>ε :92.8 % | 4.34 d | 306.86<br>545.117 | 81<br>4.3 | $^{102}$Pd(d,2pn)<br>$^{104}$Pd(d,2p3n)<br>$^{105}$Pd(d,2p4n)<br>$^{106}$Pd(d,2p5n)<br>$^{108}$Pd(d,2p7n)<br>$^{101}$Pd decay | -10033.7<br>-27640.3<br>-34734.4<br>-44295.4<br>-60055.0<br>-8691.4 |
| **$^{101g}$Rh**<br>ε: 100 % | 3.3 y | 127.23<br>198.01<br>325.23 | 68<br>73<br>11.8 | $^{102}$Pd(d,2pn)<br>$^{104}$Pd(d,2p3n)<br>$^{105}$Pd(d,2p4n)<br>$^{106}$Pd(d,2p5n)<br>$^{108}$Pd(d,2p7n)<br>$^{101}$Pd decay | -10033.7<br>-27640.3<br>-34734.4<br>-44295.4<br>-60055.0<br>-8691.38 |
| **$^{100}$Rh**<br>ε: 100 % | 20.8 h | 446.15<br>539.51<br>822.65<br>1107.22<br>1553.35 | 11.98<br>80.6<br>21.09<br>13.57<br>20.67 | $^{102}$Pd(d,2p2n)<br>$^{104}$Pd(d,2p4n)<br>$^{105}$Pd(d,2p5n)<br>$^{106}$Pd(d,2p6n)<br>$^{108}$Pd(d,2p8n)<br>$^{100}$Pd decay | -19927.6<br>-37534.3<br>-44628.4<br>-54189.3<br>-69948.9<br>-21071.0 |
| **$^{99m}$Rh**<br>64.3 keV<br>ε: 100 %<br>$\beta^+$ 7.3 % | 4.7 h | 340.8<br>617.8<br>1261.2 | 69<br>11.8<br>10.9 | $^{102}$Pd(d,2p3n)<br>$^{104}$Pd(d,2p5n)<br>$^{105}$Pd(d,2p6n)<br>$^{106}$Pd(d,2p7n)<br>$^{99}$Pd decay | -28009.2<br>-45615.8<br>-52709.9<br>-62270.9<br>-32188.3 |
| $^{99g}$Rh<br>β+: 100 % | 16.1 d | 89.76<br>353.05<br>528.24 | 33.4<br>34.6<br>38.0 | $^{102}$Pd(d,2p3n)<br>$^{104}$Pd(d,2p5n)<br>$^{105}$Pd(d,2p6n)<br>$^{106}$Pd(d,2p7n) | -28009.2<br>-45615.8<br>-52709.9<br>-62270.9 |
| **$^{103}$Ru**<br>β-: 100 % | 39.247 d | 497.09 | 91.0 | $^{105}$Pd(d,3pn)<br>$^{106}$Pd(d,3p2n) | -17956.6<br>-27517.6 |
| | | | | $^{108}$Pd(d,3p4n)<br>$^{110}$Pd(d,3p6n) | -43277.2<br>-58227.0 |
| **$^{97}$Ru**<br>ε: 100 % | 2.83 d | 215.70<br>324.49 | 85.62<br>10.79 | $^{102}$Pd(d,3p4n)<br>$^{104}$Pd(d,3p6n)<br>$^{105}$Pd(d,3p7n)<br>$^{97}$Rh decay | -42828.2<br>-60434.8<br>-67528.9<br>-47133.5 |

Increase the Q-values if compound particles are emitted by: np-d, +2.2 MeV; 2np-t, +8.48 MeV; n2p-$^{3}$He, +7.72 MeV; 2n2p-α, +28.30 MeV.

Decrease Q-values for isomeric states with level energy of the isomer

Abundances in $^{nat}$Pd(%): $^{102}$Pd(1.02), $^{104}$Pd(11.14), $^{105}$Pd(22.33), $^{106}$Pd(27.33), $^{108}$Pd(26.46), $^{110}$Pd(11.72)

# 3. Comparison with nuclear model calculations

Updated versions of ALICE-IPPE (Dityuk et al., 1998) and EMPIRE (Herman et al., 2007) codes were used to analyze the present experimental results. As described in more detail in (Hermanne et al., 2009; Ignatyuk, 2010, 2011; Tárkányi et al., 2007) these codes are named ALICE-D and EMPIRE-D and were developed by taking into account direct (d,p) and (d,t) transitions with the general relations for a nucleon transfer probability in the continuum through an energy dependent enhancement factor for the corresponding transitions. For calculation for separate levels with the ALICE-D codes, the isomeric ratios derived with EMPIRE s were applied to the ALICE-D reaction cross-sections.

The cross-sections of the investigated reactions were compared with the data given in the on-line TENDL-2010 (Koning and Rochman, 2010), -2014 (Koning et al., 2014), -2015 (Koning et al., 2015) and in case of silver products the TENDL-2009 libraries (to illustrate the development). These libraries are based on both default and adjusted parameters in TALYS (1.4) and TALYS (1.6) calculations (Koning and Rochman, 2012).

# 4. Results

## Cross-sections

The present experimental data and the theoretical description of the different reaction products are discussed briefly in separate sections. The obtained experimental data are shown in Figures 1-20 and the numerical values are tabulated in Table 2.

Table 2 Numerical data m of the measured experimental data

| E | ΔE | σ | Δσ | σ | Δσ | σ | Δσ | σ | Δσ |
|---|---|---|---|---|---|---|---|---|---|
| MeV | | mb | | | | | | | |
| | | $^{111}$Ag | | $^{110m}$Ag | | $^{106m}$Ag | | $^{105}$Ag | |
| 48.63 | 0.2 | 5.38 | 1.25 | | | 70.70 | 8.04 | 162.81 | 18.30 |
| 46.59 | 0.29 | | | 2.50 | 0.59 | 81.56 | 9.25 | 160.01 | 18.00 |
| 44.48 | 0.39 | | | | | 89.71 | 10.18 | 148.92 | 16.74 |
| 42.28 | 0.49 | | | 2.83 | 0.38 | 104.89 | 11.87 | 140.05 | 15.75 |
| 40.00 | 0.59 | | | | | 121.54 | 13.73 | 133.94 | 15.07 |
| 37.61 | 0.70 | | | 3.93 | 0.52 | 111.01 | 12.55 | 140.12 | 15.76 |
| 35.11 | 0.81 | | | 3.58 | 0.48 | 103.29 | 11.69 | 161.83 | 18.19 |
| 33.30 | 0.89 | | | 3.58 | 0.45 | 87.07 | 9.86 | 196.22 | 22.03 |
| 31.43 | 0.97 | 8.92 | 2.13 | 4.67 | 0.83 | 63.18 | 7.19 | 229.02 | 25.71 |
| | | $^{104}$Ag | | $^{103}$Ag | | $^{111m}$Pd | | $^{109}$Pd | |
| 48.63 | 0.2 | 89.76 | 11.51 | 113.79 | 16.48 | 0.359 | 0.115 | 50.05 | 5.78 |
| 46.59 | 0.29 | 96.50 | 12.34 | 125.65 | 17.70 | 0.450 | 0.130 | 51.41 | 5.91 |
| 44.48 | 0.39 | 117.93 | 14.07 | 106.33 | 14.32 | 0.709 | 0.177 | 51.93 | 5.99 |
| 42.28 | 0.49 | 124.98 | 14.67 | 113.43 | 14.26 | 0.704 | 0.142 | 49.84 | 5.74 |
| 40.00 | 0.59 | 134.62 | 15.53 | 105.43 | 12.76 | 0.610 | 0.138 | 50.53 | 5.75 |
| 37.61 | 0.70 | 121.70 | 14.06 | 109.18 | 13.27 | 0.883 | 0.165 | 52.55 | 6.02 |
| 35.11 | 0.81 | 121.94 | 14.28 | 100.80 | 12.54 | 0.887 | 0.157 | 47.11 | 5.43 |
| 33.30 | 0.89 | 103.88 | 12.64 | 102.07 | 13.90 | 0.918 | 0.169 | 47.07 | 5.42 |
| 31.43 | 0.97 | 113.40 | 13.05 | 68.37 | 8.29 | 0.893 | 0.115 | 47.27 | 5.44 |
| | | $^{101}$Pd | | $^{100}$Pd | | $^{105}$Rh | | $^{102m}$Rh | |
| 48.63 | 0.2 | 17.96 | 2.09 | 8.76 | 1.00 | 12.07 | 1.44 | | |
| 46.59 | 0.29 | 13.29 | 1.57 | 9.53 | 1.23 | 11.82 | 1.41 | 7.43 | 1.43 |
| 44.48 | 0.39 | 9.12 | 1.06 | 8.08 | 1.09 | 10.60 | 1.26 | | |
| 42.28 | 0.49 | 6.46 | 0.84 | 9.30 | 1.28 | 11.06 | 1.31 | 6.14 | 1.31 |
| 40.00 | 0.59 | 6.19 | 0.86 | 5.65 | 0.93 | 9.64 | 1.17 | | |
| 37.61 | 0.70 | 6.31 | 0.82 | 5.16 | 0.64 | 8.47 | 1.06 | 8.02 | 1.60 |
| 35.11 | 0.81 | 6.41 | 0.83 | 2.51 | 0.37 | 8.46 | 1.07 | 6.20 | 1.45 |
| 33.30 | 0.89 | 7.02 | 0.89 | 1.52 | 0.46 | 7.20 | 0.88 | 7.27 | 0.97 |
| 31.43 | 0.97 | 6.62 | 0.83 | 2.25 | 0.35 | 6.89 | 0.84 | 7.03 | 0.94 |
| | | $^{102g}$Rh | | $^{101m}$Rh | | $^{101g}$Rh | | $^{100}$Rh | |
| 48.63 | 0.2 | | | 43.00 | 4.83 | | | 24.48 | 2.76 |
| 46.59 | 0.29 | 10.55 | 5.58 | 37.28 | 4.19 | 6.77 | 1.02 | 23.84 | 2.68 |
| 44.48 | 0.39 | | | 30.27 | 3.40 | | | 21.29 | 2.39 |
| 42.28 | 0.49 | 8.76 | 3.35 | 27.97 | 3.15 | 6.94 | 1.02 | 19.34 | 2.17 |
| 40.00 | 0.59 | | | 24.97 | 2.81 | | | 17.16 | 1.93 |
| 37.61 | 0.70 | 11.35 | 1.97 | 23.01 | 2.58 | 4.54 | 0.88 | 12.91 | 1.46 |
| 35.11 | 0.81 | 8.93 | 2.25 | 21.74 | 2.44 | 3.87 | 0.72 | 9.36 | 1.06 |
| 33.30 | 0.89 | 10.21 | 3.90 | 20.67 | 2.33 | 4.51 | 0.74 | 7.11 | 0.80 |
| 31.43 | 0.97 | 10.08 | 4.07 | 20.13 | 2.26 | 3.65 | 0.47 | 5.15 | 0.58 |
| | | $^{99m}$Rh | | $^{99g}$Rh | | $^{103}$Ru | | $^{97}$Ru | |
| 48.63 | 0.2 | 10.75 | 1.22 | 3.06 | 0.42 | 0.89 | 0.13 | 1.16 | 0.13 |
| 46.59 | 0.29 | 8.40 | 0.95 | 2.96 | 0.46 | 0.67 | 0.16 | 1.08 | 0.14 |
| 44.48 | 0.39 | 6.26 | 0.71 | 2.14 | 0.36 | | | 1.02 | 0.14 |
| 42.28 | 0.49 | 4.83 | 0.55 | 1.46 | 0.31 | | | 0.53 | 0.11 |

| E | ΔE | σ | Δσ | σ | Δσ | σ | Δσ | σ | Δσ |
|---|---|---|---|---|---|---|---|---|---|
| MeV | | mb | | | | | | | |
| 40.00 | 0.59 | 2.99 | 0.35 | 0.85 | 0.2715 | 0.3205 | 0.12 | 0.28 | 0.21 |
| 37.61 | 0.70 | 1.56 | 0.19 | | | | | 0.12 | 0.03 |
| 35.11 | 0.81 | 0.70 | 0.10 | | | | | | |

## 4. 1 Production of silver radioisotopes

### *4.1.1 Cross-sections for the $^{nat}$Pd(d,x)$^{111}$Ag reaction*

The radionuclide $^{111}$Ag has two longer-lived states: a ground-state ($T_{1/2}$ = 7.45 d), and an excited isomeric state $^{111m}$Ag ($T_{1/2}$ = 64.8 s) decaying by internal transition (99.3%) to $^{111g}$Ag. The metastable and ground-state of $^{111}$Ag are also populated by the $\beta^-$decay of the metastable and ground-states $^{111}$Pd (half-lives of 5.5 h and 23.4 min). The present values include the full cumulative production including contributions of all parents. The present high energy data are consistent with the earlier experimental data (Fig. 1). The data from the three TENDL libraries significantly underestimate the experiment, while ALICE-D gives the best approximation over the relevant energy range.

### *4.1.2 Cross-sections for the $^{nat}$Pd(d,x)$^{110m}$Ag reaction*

The radionuclide $^{110}$Ag has a short-lived ground-state ($T_{1/2}$ = 24.5 s) and an excited isomeric state$^{110m}$Ag ($T_{1/2}$ = 249.8 d) that decays by internal transition to $^{110g}$Ag (IT = 1.4 %) and by $\beta^-$ decay to stable $^{110}$Cd (98.6 %). We could measure only the production of the long-lived isomer. The agreement with the earlier experimental data is good (Fig. 2). All theoretical codes overestimate the experimental data below 14 MeV, and behave differently above this energy.

### *4.1.3 Cross-sections for the $^{nat}$Pd(d,x)$^{106m}$Ag reaction*

The radionuclide $^{106}$Ag has a short-lived ground-state ($T_{1/2}$ =23.96 min) and an excited isomeric state $^{106m}$Ag ($T_{1/2}$ = 8.28 d, decay by EC (100 %) to stable $^{106}$Pd). The rather long waiting time after EOB, did not allow assessing the activity of the short-lived ground-state hence, only results for the longer-lived isomeric state are given (Fig. 3). The theory overestimates the experiment in the whole energy range.

### 4.1.4 Cross-sections for the $^{nat}$Pd(d,x) $^{105}$Ag reaction

The radionuclide $^{105}$Ag has a ground-state ($T_{1/2}$ = 41.29 d) and a short-lived excited isomeric state $^{105m}$Ag ($T_{1/2}$ = 7.23 min) that decays by internal transition (99.66 %) to $^{105}$Ag. We could only measure the cumulative production of $^{105g}$Ag (i.e. m+). The present data agree well with our earlier values and the TENDL approximations are close to the experimental data (Fig. 4).

### 4.1.5 Cross-sections for the $^{nat}$Pd(d,x) $^{104}$Ag reaction

The radionuclide $^{104}$Ag also has a short-lived ground-state ($T_{1/2}$ = 69.2 min), and a shorter-lived excited isomeric state ($T_{1/2}$= 33.5 min) that decays by internal transition (0.07 %) to $^{104g}$Ag. In the first sets of spectra used for determination of the cross-sections of $^{104}$Ag, the cooling time was around 9 hours and we could not detect independent gamma-lines from decay of $^{104m}$Ag, while the stronger 550 keV common gamma-line gives cross-sections identical to the cross-sections obtained from independent gamma-lines of $^{104g}$Ag. Accordingly, the measured cross-sections for $^{104g}$Ag are cumulative (m+), in other words the $^{104m}$Ag contribution is negligible by the following reason: At the time of spectra measurement the ground-state had decayed for 7-8 half-lives and the metastable state for 16 half-lives resulting in an activity ratio of around 250-300. Furthermore, the low transition rate (0.0007) introduces additional factor of 1/0.0007 =1430 and the ground-state hence contains only a very small contribution of metastable state decay: total factor 1/3.5 $10^5$)

The recent theoretical predictions overestimate the experiment (Fig. 5). The importance of experimental data on code development is clearly seen: the magnitude of the TENDL prediction changed by factor of 5 in the last 7 years.

### 4.1.6 Cross-sections for the $^{nat}$Pd(d,x) $^{103}$Ag reaction

The radionuclide $^{103}$Ag has a ground-state ($T_{1/2}$ = 65.7 min), and a short-lived excited isomeric state$^{103m}$Ag ($T_{1/2}$ = 5.7 s) that decays by internal transition (100 %) to $^{103g}$Ag. We could measure only the cumulative production of $^{103g}$Ag (i.e. m+). The new versions of TENDL describe the experimental data within about 10 % (Fig. 6).

## 4. 2 Production of palladium radioisotopes

### *4.2.1 Cross-sections for the $^{nat}$Pd(d,x) $^{111m}$Pd reaction*

The radionuclide $^{111}$Pd has a short-lived ground-state ($T_{1/2}$ = 23.4 min, $\beta^-$:100 %) and a longer-lived metastable state ($T_{1/2}$ = 5.5 h) decaying with IT (73 %) to the ground-state and with $\beta^-$(23 %) to$^{111}$Ag. We obtained experimental cross-sections for the production of the metastable state (Fig. 7). The theoretical description, in particular ALICE-D show fairly good agreement with the data in the present energy range (tail of the $^{110}$Pd(d,p) excitation curve), but differs largely in the region of maximum up to a factor of 2-3.

### *4.2.2 Cross-sections for the $^{nat}$Pd(d,x)$^{109}$Pd reaction*

The higher laying metastable state ($T_{1/2}$ = 4.69 min) decays with IT (100%) to the $^{109}$Pd ground state ($T_{1/2}$ = 13.7012 h, $\beta^-$:100%). The measured cross-sections are fully cumulative. The new data show good agreement with the earlier experiments (Fig. 8). The theoretical descriptions of the TENDL libraries are in fair agreement but far from satisfactory in the low energy region.

### *4.2.3 Cross-sections for the $^{nat}$Pd(d,x) $^{101}$Pd reaction*

The $^{101}$Pd ($T_{1/2}$ = 8.47 h) radionuclide is produced directly and through decay of $^{101g}$Ag ($T_{1/2}$ = 11.1 min, EC(100 %)) including $^{101m}$Ag ($T_{1/2}$ = 3.1 s, IT 100 %). The measured production cross-sections of $^{101}$Pd are cumulative, obtained from the spectra measured after ‘‘complete’’ decay of $^{101}$Ag (Fig. 9). The contributions of the reactions on $^{102}$Pd and $^{104}$Pd can be distinguished in experimental and theoretical values (two maxima). EMPIRE-D gives fairly good agreement with our values.

### *4.2.4 Cross-sections for the $^{nat}$Pd(d,x) $^{100}$Pd reaction*

The radionuclide $^{100}$Pd ($T_{1/2}$ = 3.36 d) is also formed from the EC + $\beta^+$decay of the short-lived ground and first isomeric states of $^{100}$Ag ($^{100g}$Ag: ($T_{1/2}$ = 2.01 min) and $^{100m}$Ag: ($T_{1/2}$ = 2.24 min)). Cross-sections were deduced from spectra measured after total decay of the $^{100}$Ag, hence the cross-sections are cumulative. The experimental data are discrepant below about 35 MeV, and the theoretical descriptions of the presently measured cross sections show also discrepancy (Fig. 10) although the shape of energy dependence is similar. The threshold of the $^{102}$Pd(d,tn)$^{100}$Pd reaction is lower, 12.8 MeV, but the cross sections are negligible, and the threshold of the reaction for $^{100}$Pd ($^{102}$Pd(d,4n)$^{100}$Ag) is 29.5 MeV.

## 4. 3 Production of rhodium radioisotopes

### *4.3.1 Cross-sections for the $^{nat}$Pd(d,x) $^{105}$Rh reaction*

The radionuclide $^{105}$Rh has a long-lived ground-state ($T_{1/2}$ = 35.36 h, $\beta^-$:100%) and a short-lived metastable state ($T_{1/2}$ = 40 s) decaying with IT (100%) to the ground state. Our results, measured after complete decay of the isomeric state, present cumulative cross-sections (Fig. 11). If we ignore the small contribution from the decay of the $^{105}$Ru isotope ($T_{1/2}$ = 4.44 h) produced via low cross sections (d,3pxn) reactions with threshold above 13 MeV, the $^{105}$Rh is produced only directly. The experimental data show slight difference, but the TENDL predictions are about half of the experimental values and do not reproduce the pronounced structure predicted by ALICE-D and EMPIRE-D.

### *4.3.2 Cross-sections for the $^{nat}$Pd(d,x) $^{102m}$Rh and $^{103}$Rh(p,pn)$^{102g}$Rh reactions*

The $^{102}$Rh has two long-lived isomeric states: the long half-life $^{102m}$Rh ($T_{1/2}$ = 3.742 y) and the somewhat shorter-lived ground state ($T_{1/2}$ = 207.3 d). Due to the long half-lives and the very low isomeric transition ratio (IT: 0.233 %) we could deduce independent formation cross-sections for both states. The formations are direct, the radionuclide $^{102}$Rh is a “closed nucleus” from the point of view of parent decays. The results are given in Figs. 12 and 13 in comparison with the theoretical results and with the previous results for $^{102g}$Rh by (Ditrói et al., 2012). In the case of $^{102m}$Rh no earlier experimental results were found.

### *4.3.3 Cross-sections for the $^{nat}Pd(d,x)\,^{101m}Rh$ and $^{nat}Pd(d,x)\,^{101g}Rh$ reactions*

We could identify the decay of both isomeric states of $^{101}$Rh ($^{101g}$Rh- $T_{1/2}$ = 3.3 y, $^{101m}$Rh-$T_{1/2}$ = 4.34 d). The metastable $^{101m}$Rh decays with IT (7.20 %) to the long-lived ground-state, resulting in low contribution to activity, and by EC (93.6 %) to stable $^{101}$Ru. The radionuclide $^{101m}$Rh is produced directly and by the decay of $^{101}$Pd ($T_{1/2}$ = 8.47 h, ε: 99.75 %). The ground-state is produced directly, from the decay of the isomeric state and from the decay of $^{101}$Pd (0.25 %). The cumulative cross-sections of $^{101m}$Rh were determined from the spectra taken after the "complete" decay of $^{101}$Pd, the cumulative cross-sections of $^{101g}$Rh from spectra after complete decay of $^{101}$Pd and $^{101m}$Rh (Figs. 14 and 15). For both cases, the EMPIRE-D shows marked difference from the present data while the ALICE-D is in faire agreement both in shape and magnitude.

### *4.3.4 Cross-sections for the $^{nat}Pd(d,x)^{100}Rh$ reaction*

The radionuclide $^{100}$Rh has a short-lived ($T_{1/2}$ = 4.6 min) isomeric state decaying with IT (98.3 %) to the long-lived ground-state ($T_{1/2}$ = 20.8 h, decays by EC (100%)). Our new values are cumulative as they were measured after complete decay of the isomeric state. On the other hand, as they were obtained relatively shortly after EOB and with short measuring time, the contribution from decay of the longer-lived $^{100}$Pd ($T_{1/2}$ = 3.63 d, decay by EC (100 %), see section 2.4.4) to the measured $^{100}$Rh activity is not significant (Fig. 16). The present data are in good agreement with the older ones. The newer TENDL libraries give a good description of the experimental values wile EMPIRE-D and ALICE-D are largely overestimating.

### *4.3.5 Cross-sections for the $^{nat}Pd(d,x)\,^{99m}Rh$ and $^{nat}Pd(d,x)\,^{99g}Rh$ reactios*

These two long-lived states of $^{99}$Rh decay independently. They metastable state $^{99m}$Rh is produced directly and populated also through the decay of the short-lived $^{99}$Pd parent ($T_{1/2}$ = 21.4 min). The excitation function for cumulative production of the $^{99m}$Rh metastable state ($T_{1/2}$ = 4.7 h, IT < 0.16%, EC: 92.7 %, $\beta^+$:7.3 %), including the 97.4% decay contribution of $^{99}$Pd is shown in Fig. 17 and is well represented by TENDL-2014 and TENDL-2015.

The cumulative cross-sections of the ground state $^{99g}$Rh ($T_{1/2}$ = 16.1 d, EC: 96.4 %, $\beta^+$: 3.6 %), including the 2.6 % decay contribution from the $^{99}$Pd parent are shown in Fig. 18. In this case EMPIRE-D and ALICE-D give better description.

## 4. 4 Production of ruthenium radioisotopes

### *4.4.1 Cross-sections for the $^{nat}$Pd(d,x)$^{103}$Ru reaction*

The results on cumulative cross-sections of the $^{103}$Ru ground-state ($T_{1/2}$ = 39.247 d, $\beta^-$: 100 %), including possibly 100 % decay contribution from $^{103}$Tc parent ($T_{1/2}$ = 54.2 s) are shown in Fig. 19. TENDL calculations show much better description of the data than EMPIRE and ALICE.

### *4.4.2 Cross-sections for the $^{nat}$Pd(d,x) $^{97}$Ru reaction*

The measured experimental cross-section data of $^{97}$Ru ($T_{1/2}$ = 2.83 d) are cumulative (Fig. 20), including direct reaction and possibly indirect formation through the $^{97}$Ag ($T_{1/2}$ = 25.5 s)- $^{97}$Pd ($T_{1/2}$ = 3.1 min) $^{97m,g}$Rh ($T_{1/2}$ = 30.7 min and 46.2 min) $^{97}$Ru decay chain. The results are closer to the TENDL calculations as in the case of $^{103}$Ru production.

## Integral yields

From excitation functions obtained by spline fit to our experimental cross-section data, integral physical yields (Bonardi, 1987) were calculated and shown in Figs. 21 and 22 as a function of the energy. No earlier experimental data were found from other authors.

## 5. Production routes for applications

Proton induced nuclear reactions on palladium have applications in different fields of biological studies, nuclear medicine and labeling for industrial processes. A more detailed comparison of different production routes was done in our previous paper (Tárkányi et al., 2016). Here, production routes of these isotopes are discussed considering the results of the presently studied deuteron induced reactions.

**$^{103}$Pd (trough $^{103}$Ag)** Because of its suitable half-life and decay characteristics $^{103}$Pd (16.991 d) is clinically used in permanent brachytherapy. For complete review of production routes of the $^{103}$Pd we refer to our previous work (Tárkányi et al., 2016; Tárkányi et al., 2009). At low energy accelerators ($E_{p,max}$ = 20 MeV) direct production routes via the $^{103}$Rh(p,n), $^{103}$Rh(d,2n) and indirect route via the $^{104}$Pd(p,2n) reaction can be considered. The $^{103}$Rh(p,n) process with the highest yield is industrially used. On the other hand, the excitation function for the $^{104}$Pd(d,3n)$^{103}$Ag reaction (a precursor of therapeutic $^{103}$Pd) presented here leads us to conclude that this pathway is not an attractive alternative production route for $^{103}$Pd, because cross-sections higher than 100 mb only above 30 MeV deuteron energy, where no commercial accelerators are available.

The **$^{104g}$Ag** ($T_{1/2}$ = 69.2 min, $\beta^+$ 15 %) is a PET imaging analogue of the therapeutic radionuclide $^{111}$Ag. Production routes were discussed in (Hermanne et al., 2004a). Nearly pure $^{104}$Ag (m + g) can be produced with 15 MeV protons on highly enriched $^{104}$Pd targets. When we use the presently investigated deuteron induced reactions for $^{104}$Ag production, contamination with$^{105}$Ag ($T_{1/2}$ = 41.3 d) through the $^{104}$Pd(d,n) reaction cannot be avoided.

The **$^{105,106}$Ag** longer-lived radioisotopes of silver are widely use to study Ag metabolism, labeling of nanoparticles, diffusion in solids, etc. In thin layer activation study of palladium the $^{105}$Ag product obtained through $^{nat}$Pd(p,x)$^{105}$Ag is used. At low and medium energy accelerators $^{105}$Ag ($T_{1/2}$ = 41.29 d) can be produced via $^{105}$Pd(p,n) or $^{104}$Pd(d,n) with the proton reaction is more productive. Similarly for production of $^{106}$Ag ($T_{1/2}$ = 8.28 d) the $^{106}$Pd(p,n) and $^{106}$Pd(d,2n) reactions are available. In this mass region the yield of the (d,2n) are higher than that of a (p,n) reaction on the same target, but higher energy cyclotrons (not commercially available) are needed.

The **$^{110m}$Ag** *(*$T_{1/2}$ = **243.83 d)** was used for silver uptake studies in different biological organism, for characterization of [$^{110m}$Ag]-nanoparticles in different organisms, industrial tracing of silver in long processes, etc. A highly efficient production route, but with low specific activity, is

$^{109}$Ag(n,γ). It can be produced n.c.a. (no-carrier added) via $^{110}$Pd(p,n) and $^{110}$Pd(d,2n) reaction. When using the deuteron route $^{111}$Ag is produced simultaneously via (d,n) reaction, but it will not disturb the above mentioned tracing applications in practice. The (d,2n) route requires a higher energy cyclotron, because the maximum of (d,2n) reaction is shifted about 6 MeV to higher energies compared to the (p,n) reaction although the maximum cross-section is higher .

The **$^{111}$Ag** ($T_{1/2}$ = 7.45 d) can be n.c.a. (no-carrier added) produced in a nuclear reactor via neutron capture on enriched $^{110}$Pd, $\beta^-$ decay of $^{111}$Pd and is a potential therapeutic radionuclide (radio-synovectomy) because of the production feasibility and its favorable nuclear properties. In charged particle irradiations of Pd it can be produced only with deuterons.

The radionuclides **$^{105}$Rh** *(*$T_{1/2}$ = 35.36 h)**,** **$^{101m}$Rh** ***(***4.34 d) (also as decay product of $^{101}$Pd***),*** **$^{101}$Rh** ($T_{1/2}$ = 3.3 y) and **$^{97}$Ru** ***(***$T_{1/2}$ = 2.83 d) have been considered as potential candidates for targeted radiotherapy use. Light charged particle induced reactions on isotopes of ruthenium are advantageous for the production of these rhodium isotopes.

## 6. Summary and conclusion

Twenty excitation functions of $^{nat}Pd(d,xn)^{111,110m,106m,105,104g,103}Ag$, $^{nat}Pd(d,x)$ $^{111m,109,101,100}Pd$, $^{nat}Pd(d,x)$, $^{105,102m,102g,101m,101g,100,99m,99g}Rh$ and $^{nat}Pt(d,x)^{103,97}Ru$ nuclear reactions on Pd were measured, in the 41-50 MeV region for the first time. All the data were measured relative to well documented monitor reactions. The present results are in general consistency with our previous data in the lower energy region except for some cases. The experimental data were compared with the results of a priori model calculations performed by the EMPIRE-D, ALICE-IPPE-D and TALYS(TENDL) codes. The descriptions by theoretical calculations show that the shape and the absolute values of the excitation functions is only partly successful, in some cases still large disagreements were found despite the upgrading of the codes. The obtained experimental data provide a basis for improved model calculations and for practical applications. In applications, the deuteron induced reactions can play an important role, when a proton induced route is not possible and (d,n) or (d,p) reactions can bring a solution. In the heavier mass region using (d,2n) instead of (p,n) reactions results in higher production yields.

## Acknowledgements

This work was done in the frame of MTA-FWO (Vlaanderen) research projects. The authors acknowledge the support of research projects and of their respective institutions in providing the materials and the facilities for this work.

## Figures

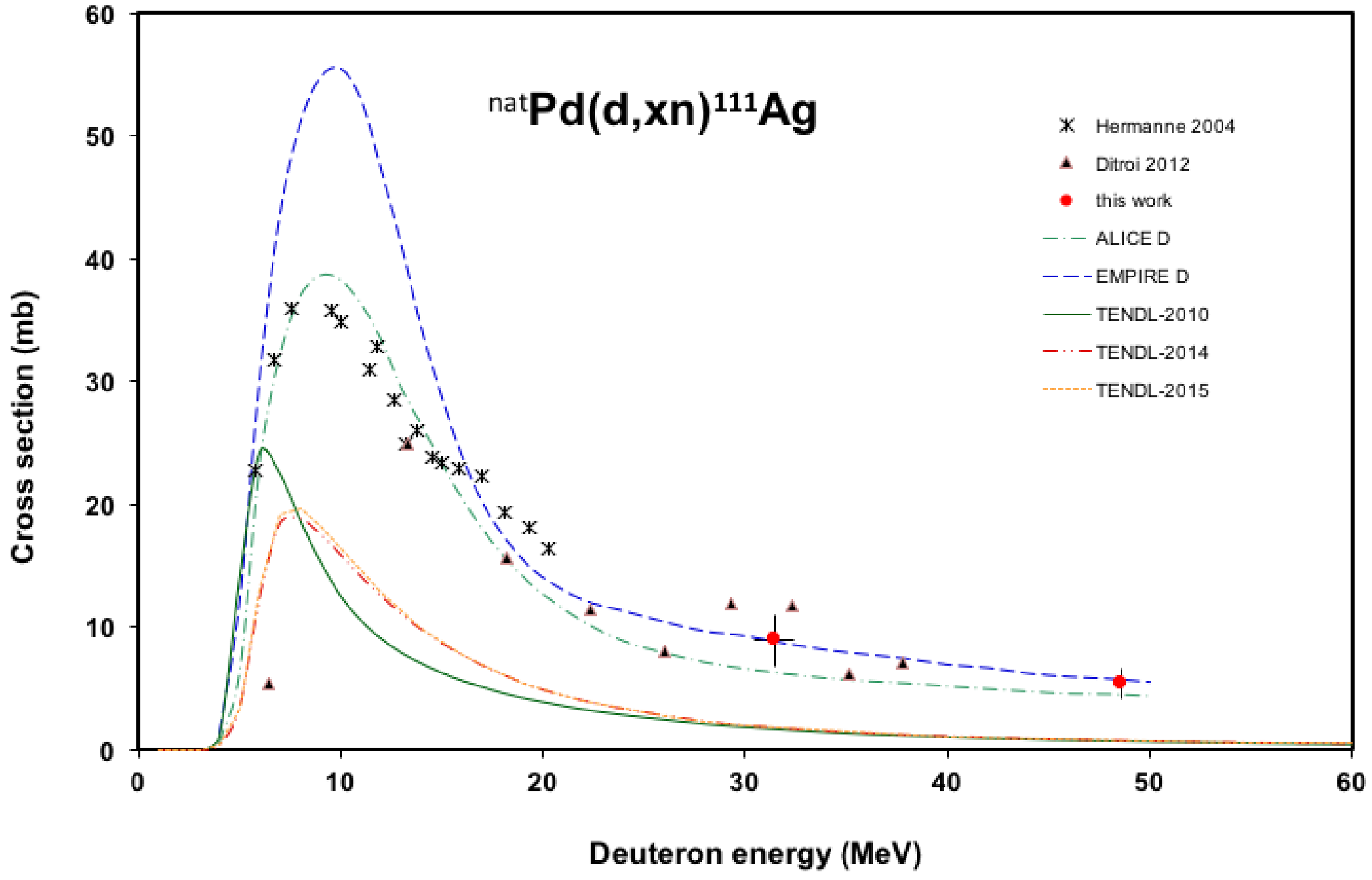

Fig. 1. Excitation functions of the $^{nat}Pd(d,x)^{111}Ag$ reaction in comparison with literature values and theoretical results

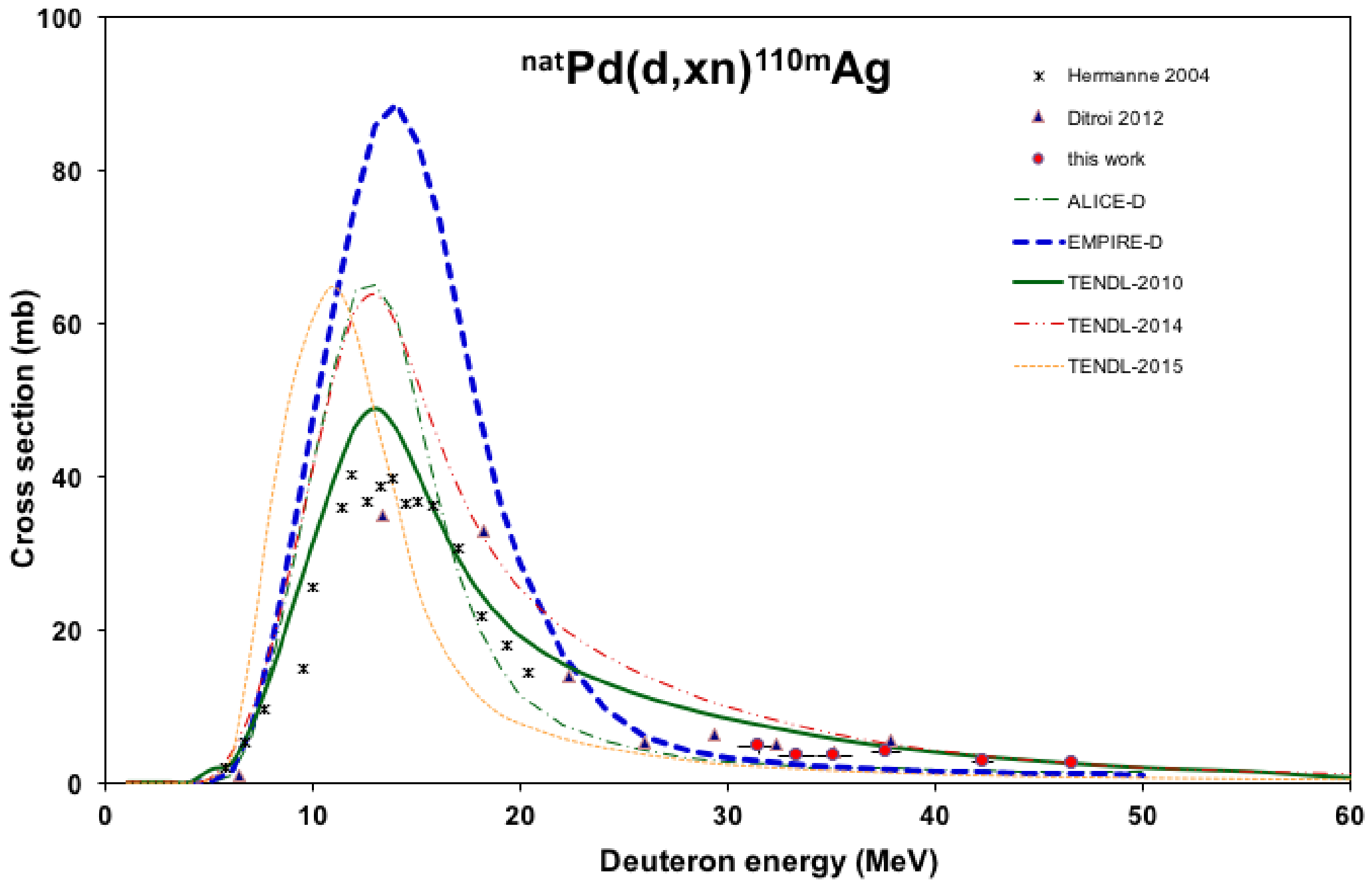

Fig. 2. Excitation functions of the $^{nat}Pd(d,x)^{110m}Ag$ reaction in comparison with literature values and theoretical results

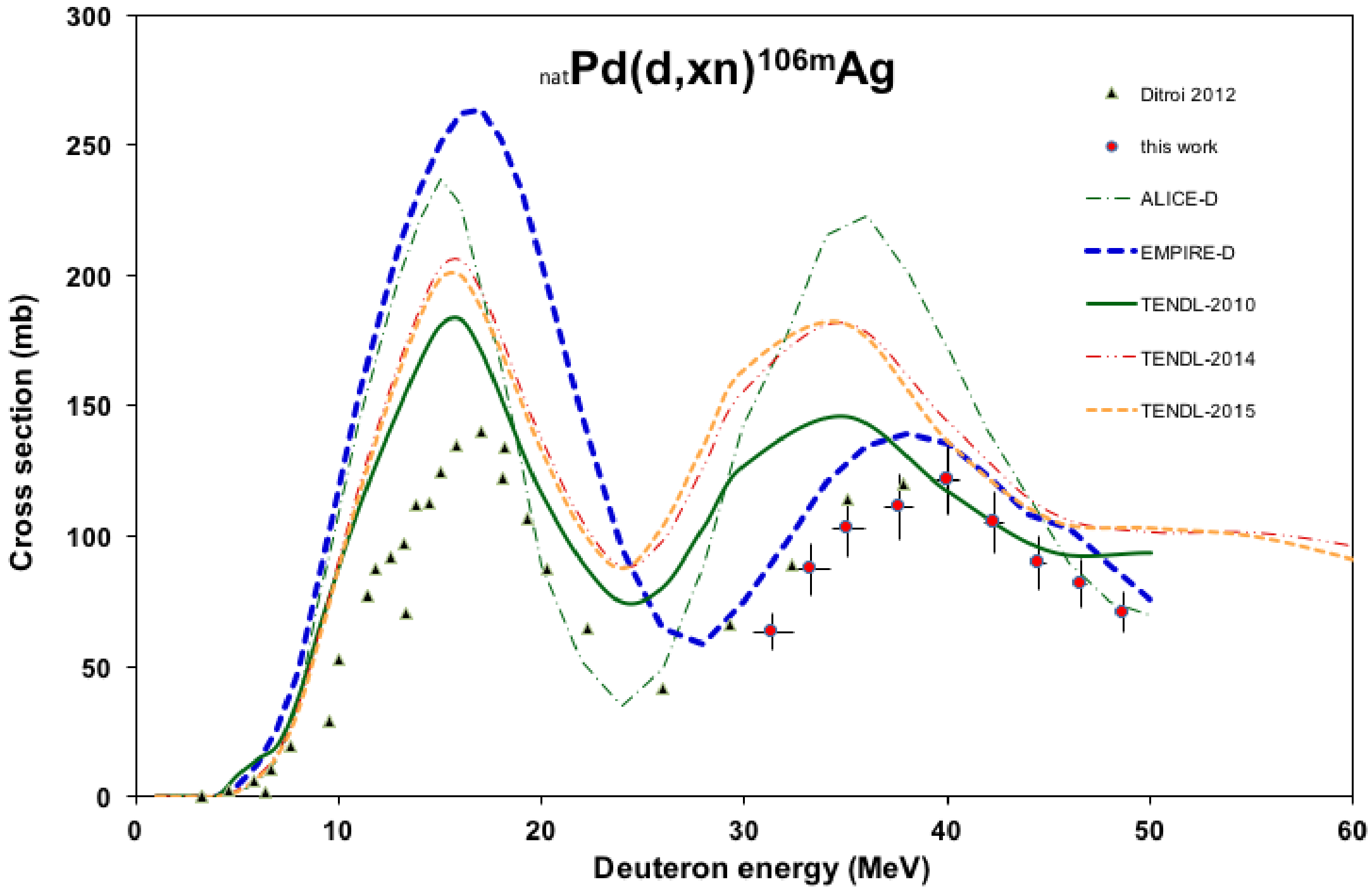

Fig. 3. Excitation functions of the $^{nat}Pd(d,x)^{106m}Ag$ reaction in comparison with literature values and theoretical results

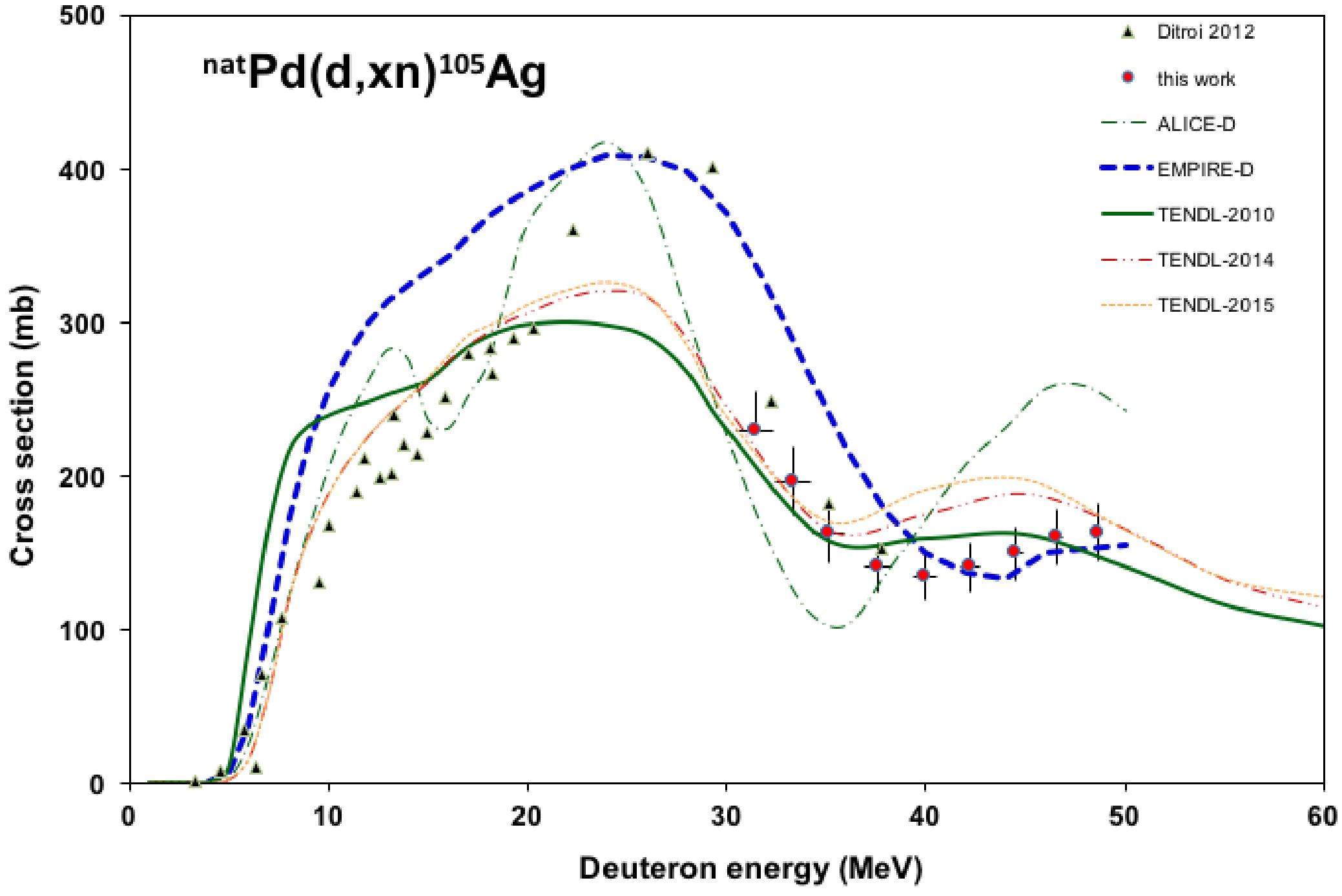

Fig. 4. Excitation functions of the $^{nat}Pd(d,x)^{105}Ag$ reaction in comparison with literature values and theoretical results

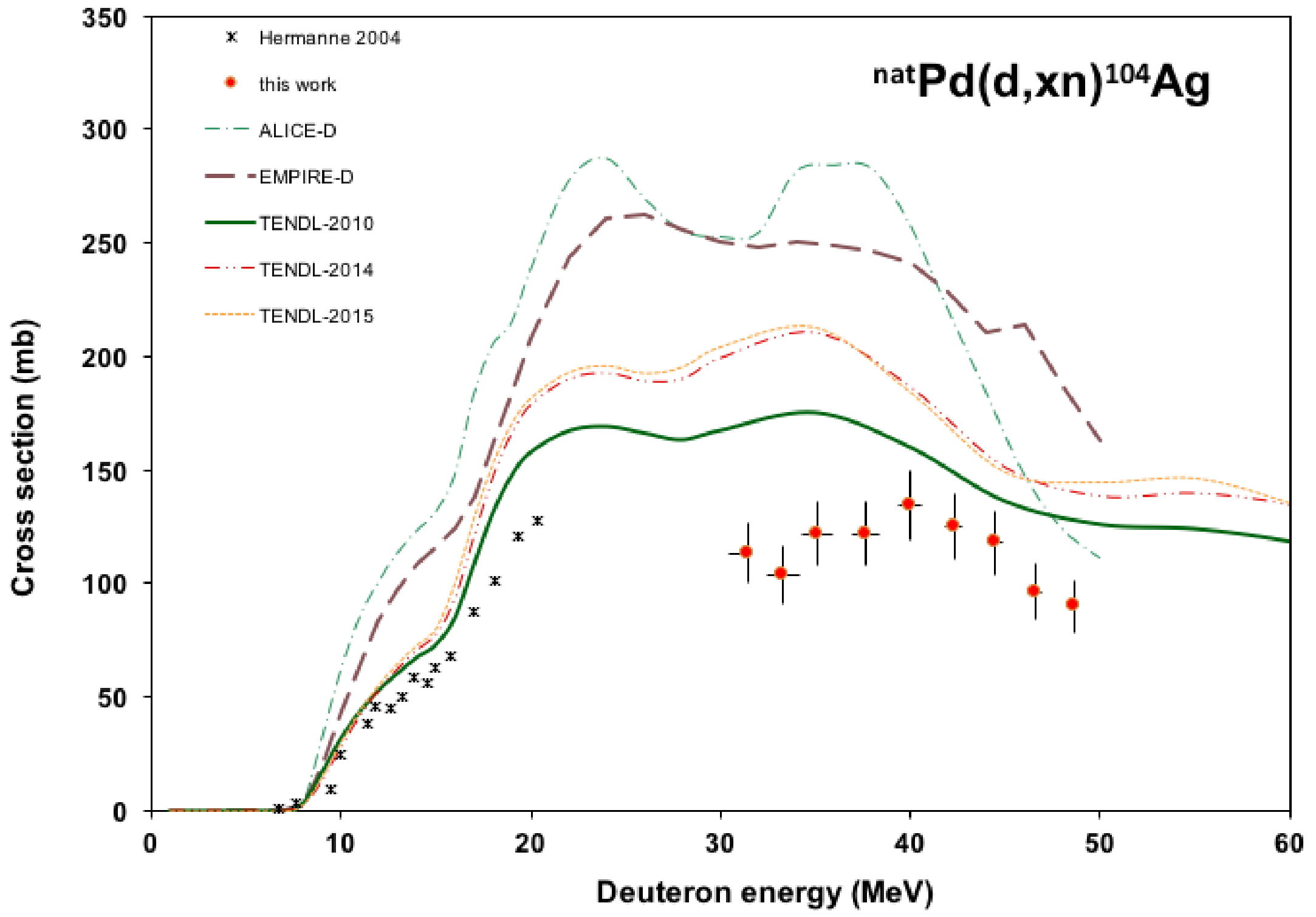

Fig. 5. Excitation functions of the $^{nat}Pd(d,x)^{104}Ag$ reaction in comparison with literature values and theoretical results

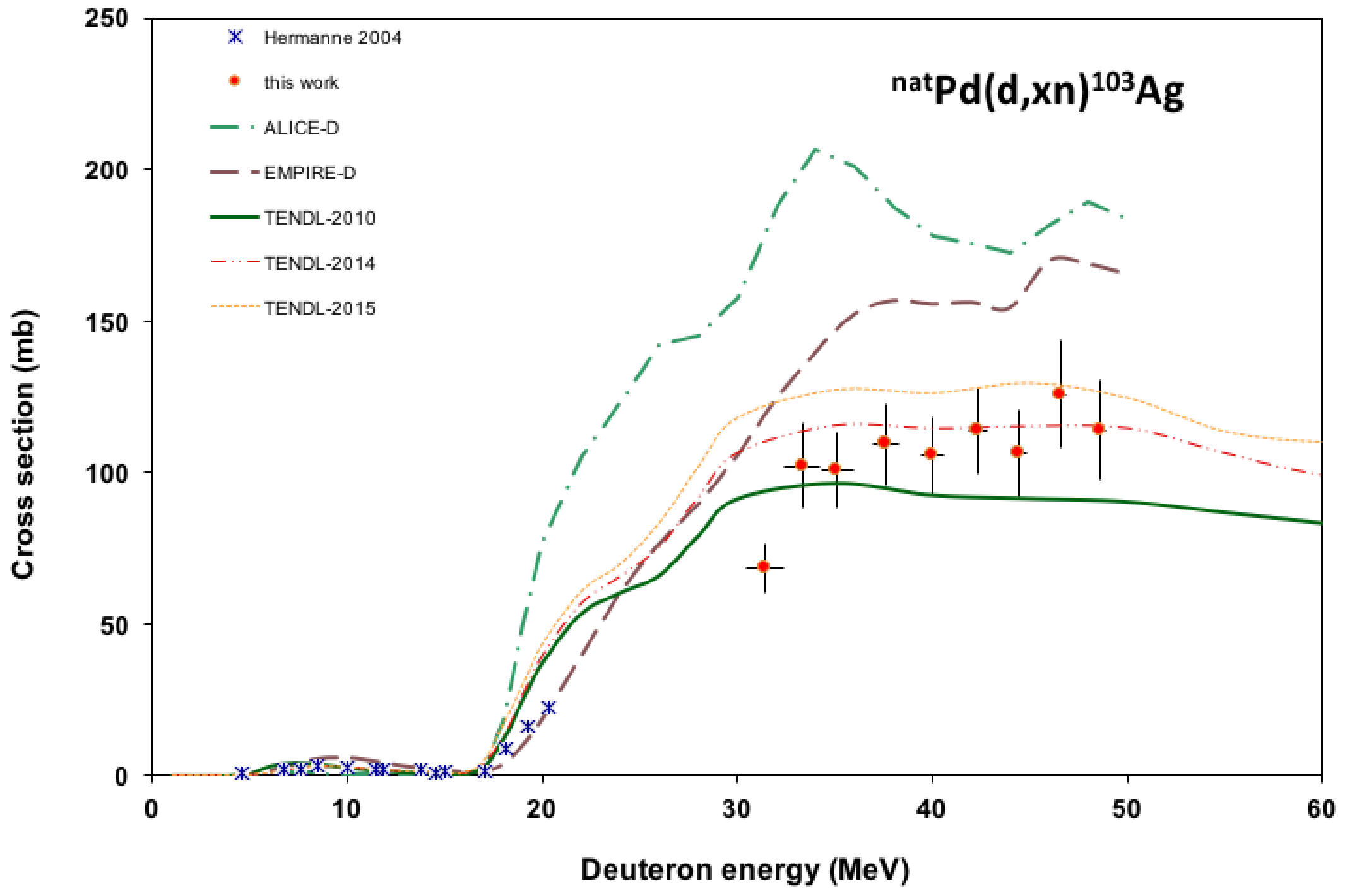

Fig. 6. Excitation functions of the $^{nat}Pd(d,x)^{103}Ag$ reaction in comparison with literature values and theoretical results

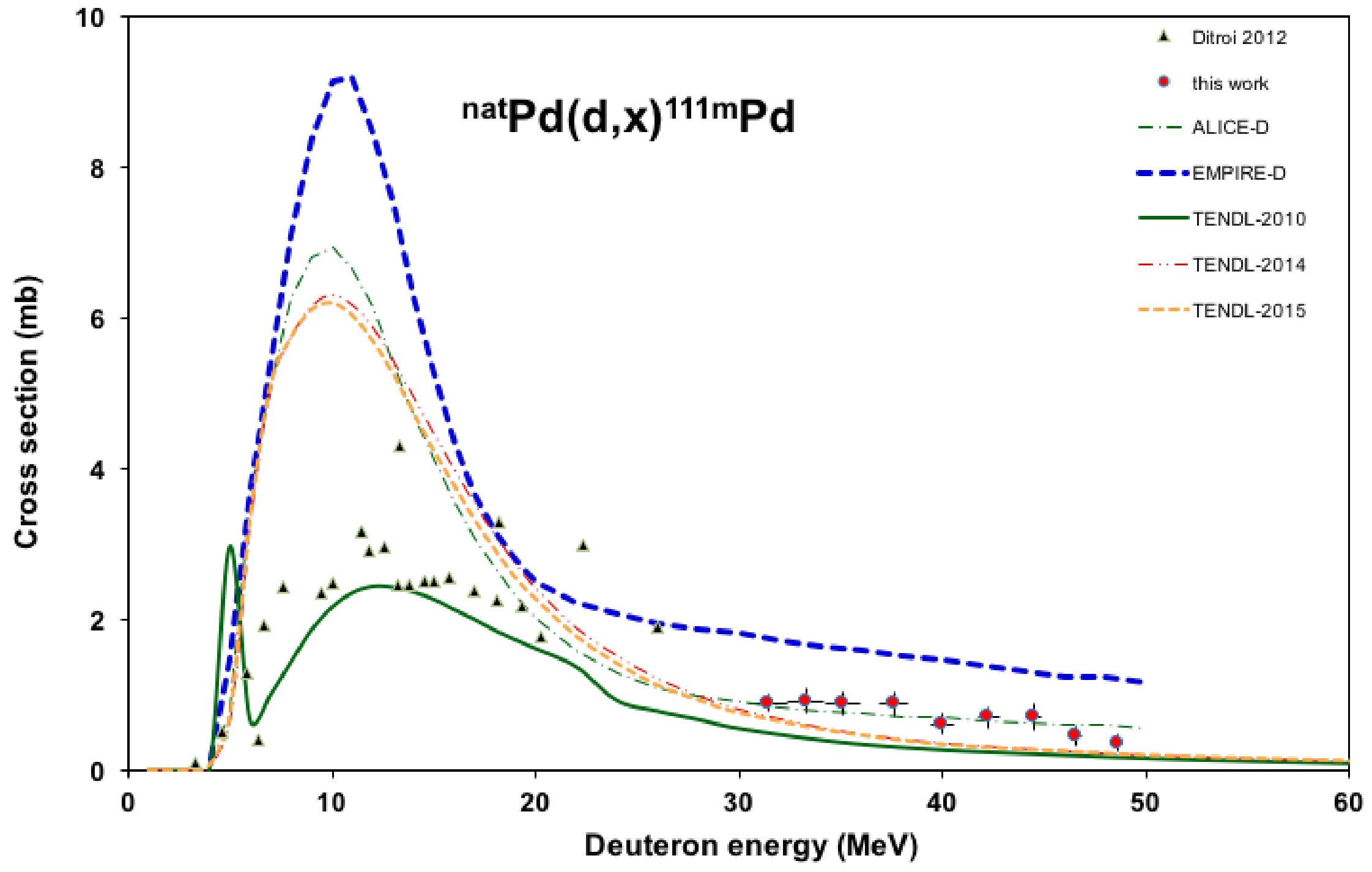

Fig. 7. Excitation functions of the $^{nat}Pd(d,x)^{111m}Pd$ reaction in comparison with literature values and theoretical results

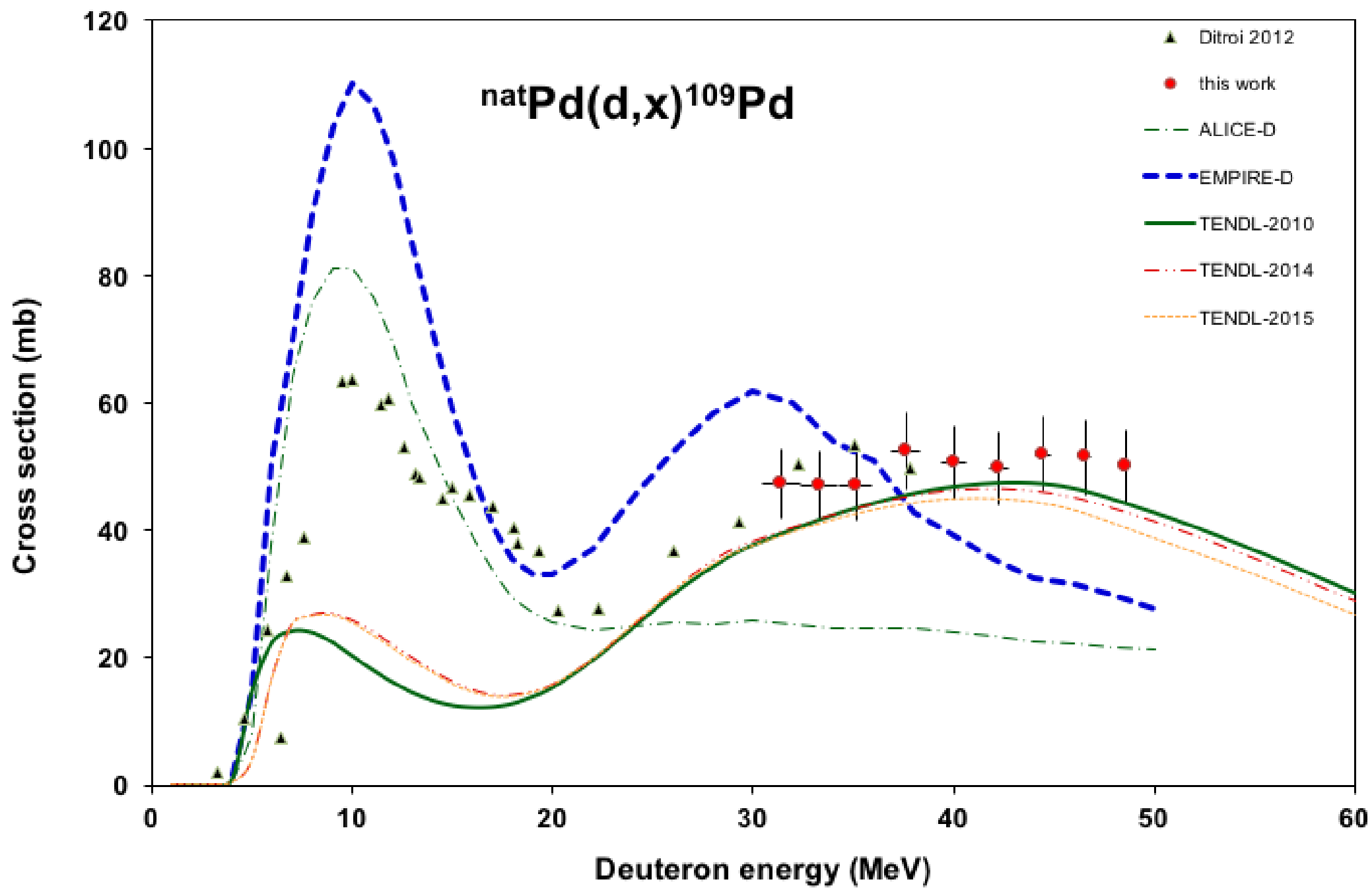

Fig. 8. Excitation functions of the $^{nat}Pd(d,x)^{109}Pd$ reaction in comparison with literature values and theoretical results

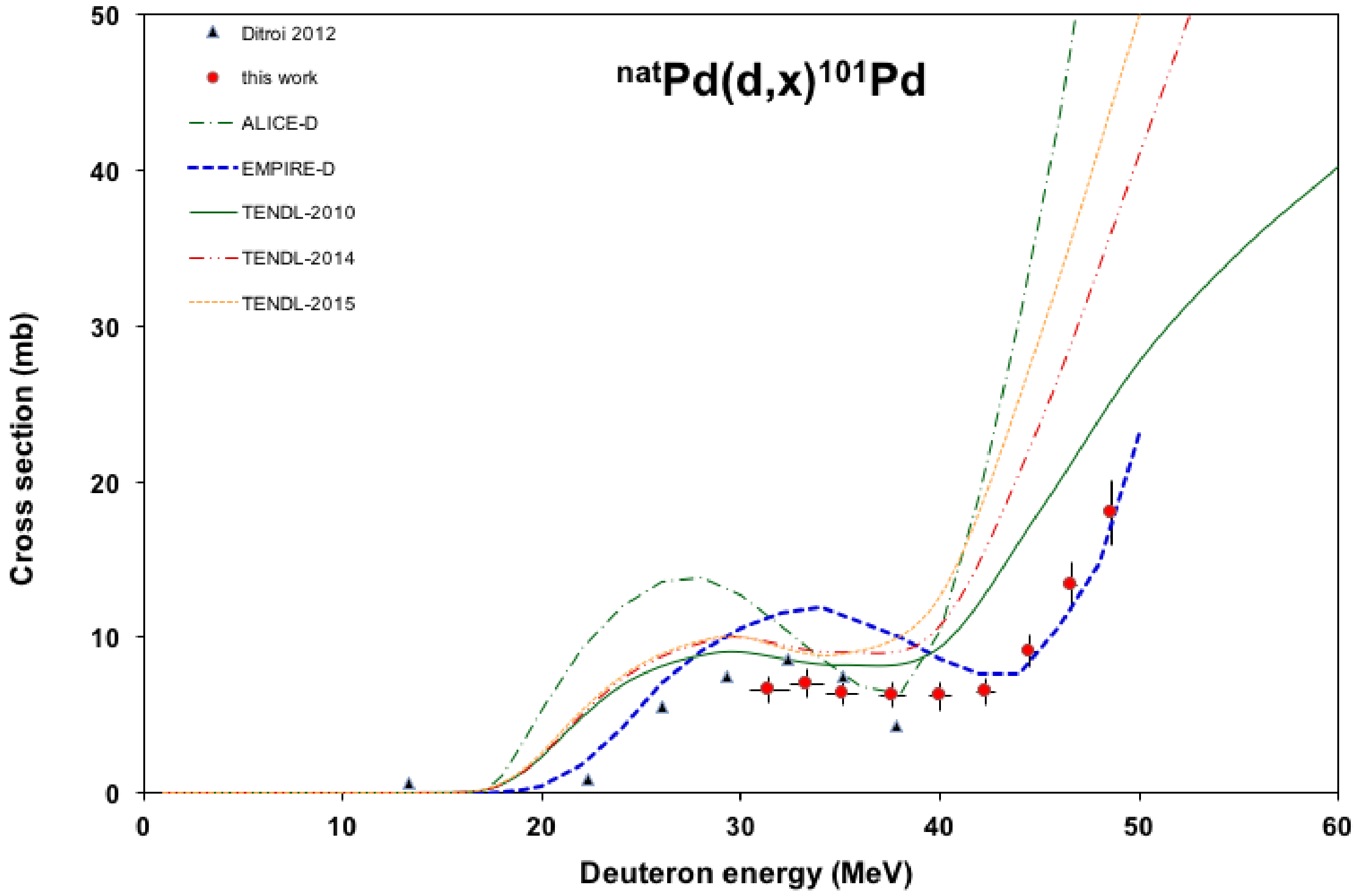

Fig. 9. Excitation functions of the $^{nat}Pd(d,x)^{101}Pd$ reaction in comparison with literature values and theoretical results

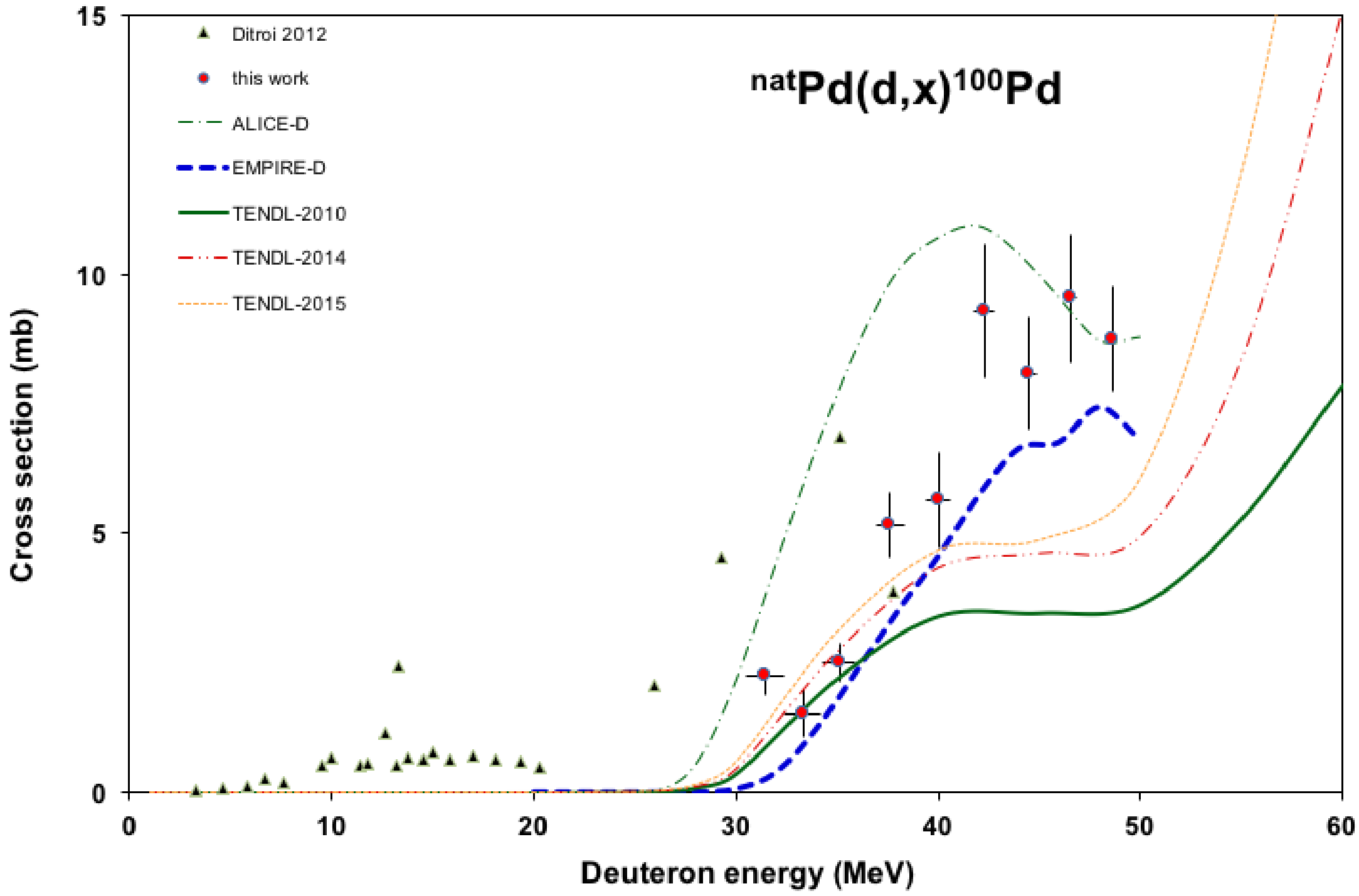

Fig. 10. Excitation functions of the $^{nat}Pd(d,x)^{100}Pd$ reaction in comparison with literature values and theoretical results

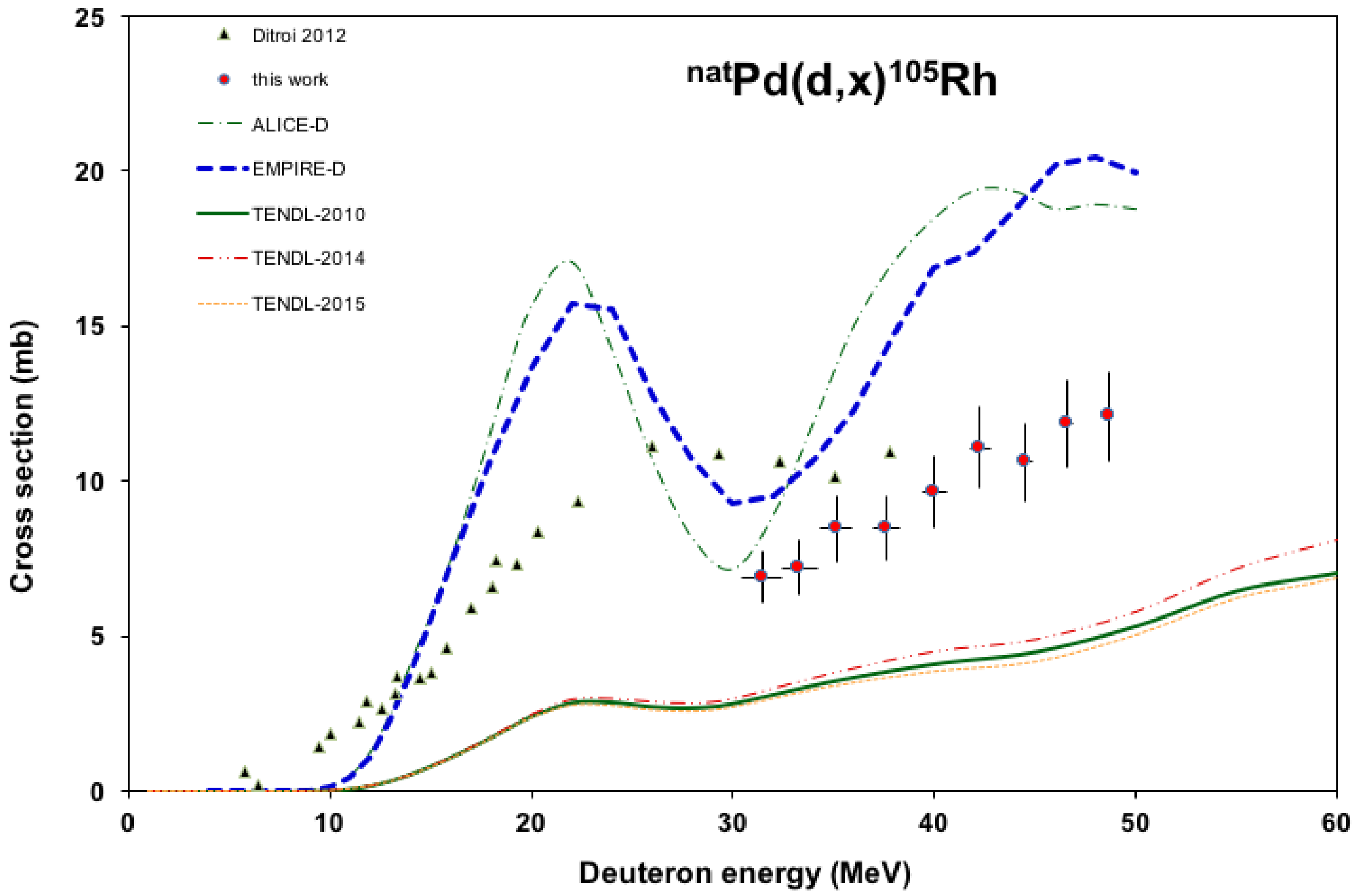

Fig. 11. Excitation functions of the $^{nat}Pd(d,x)^{105}Rh$ reaction in comparison with literature values and theoretical results

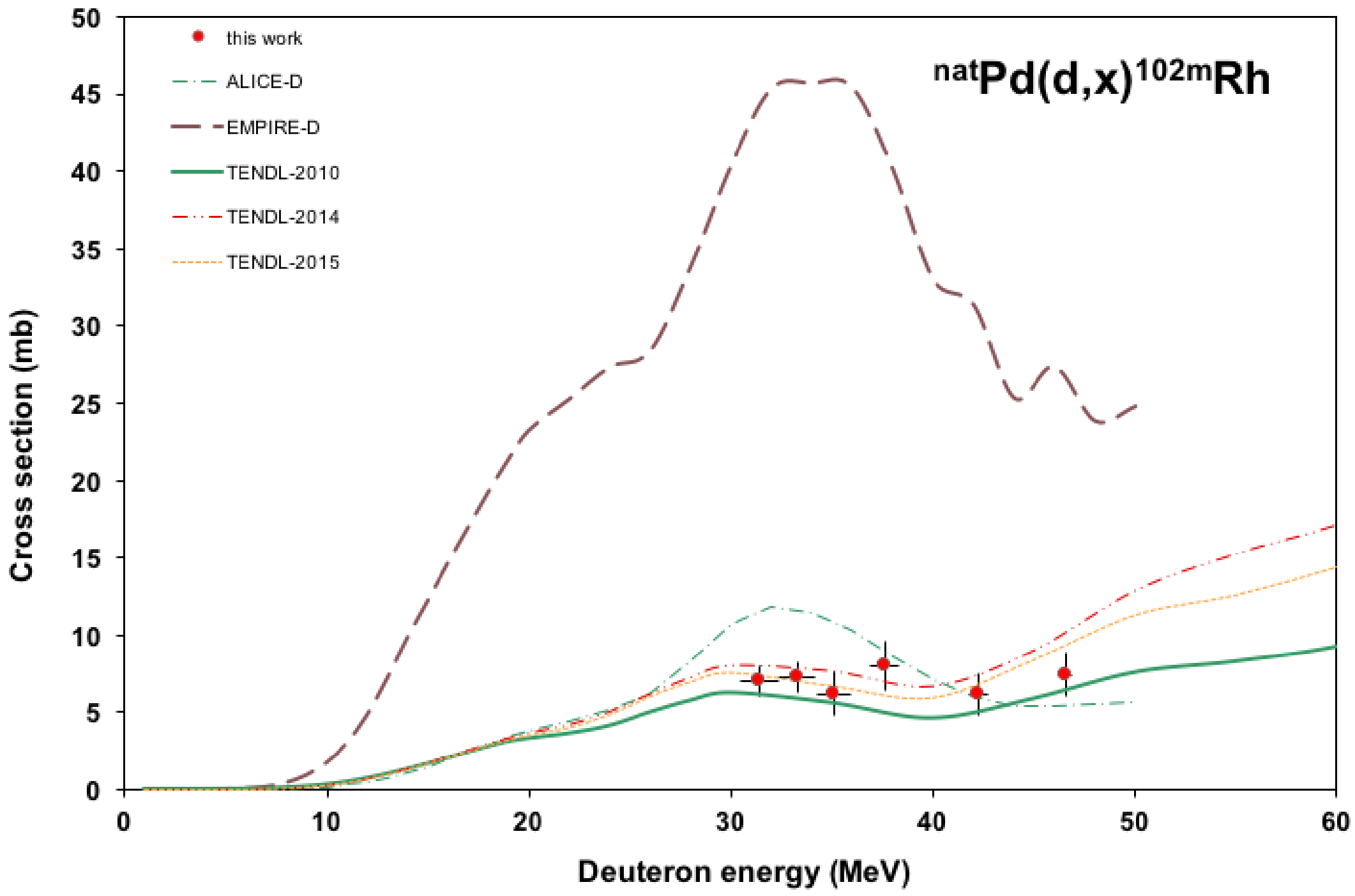

Fig. 12. Excitation functions of the $^{nat}Pd(d,x)^{102m}Rh$ reaction in comparison with literature values and theoretical results

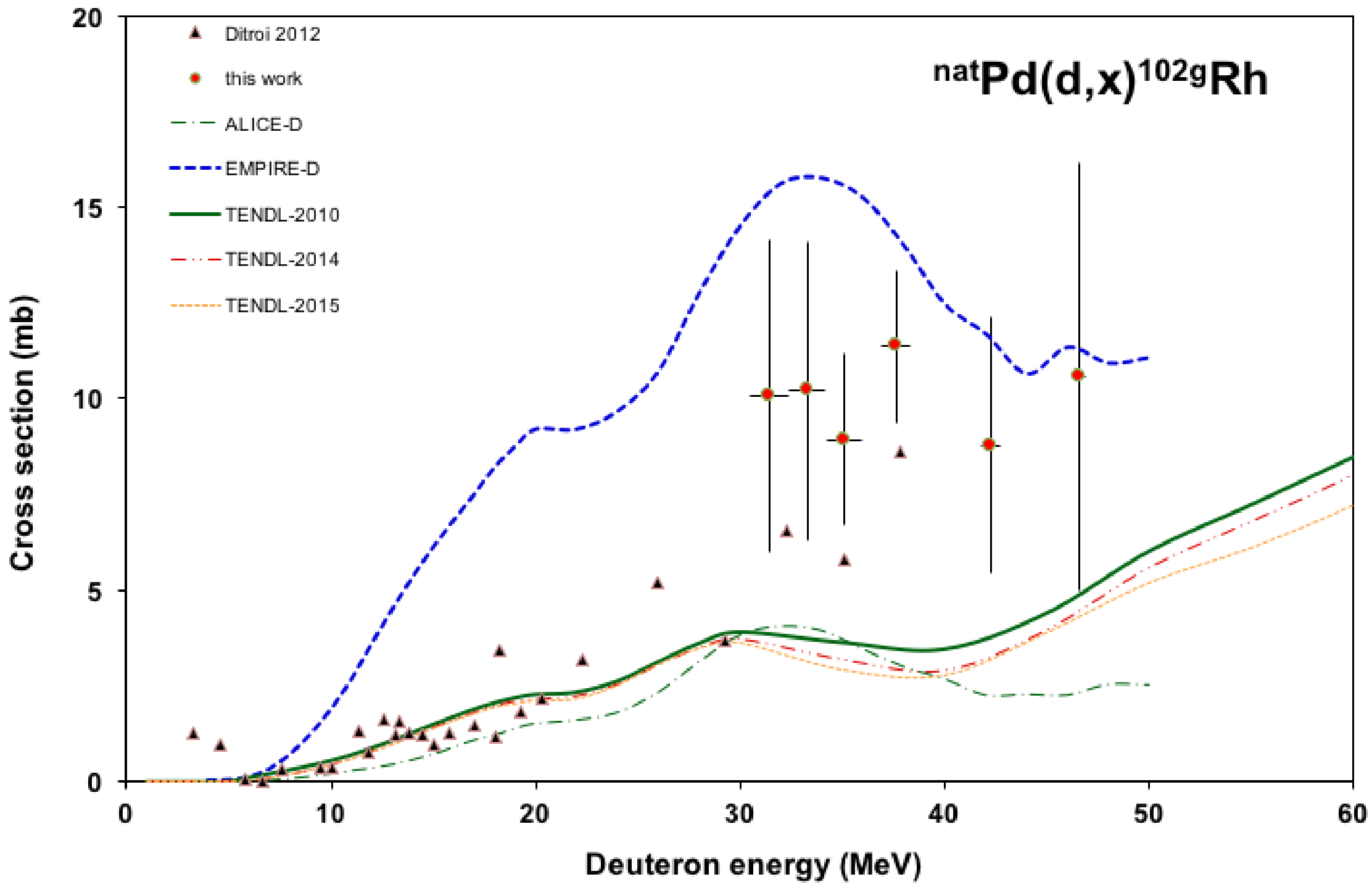

Fig. 13. Excitation functions of the $^{nat}Pd(d,x)^{102g}Rh$ reaction in comparison with literature values and theoretical results

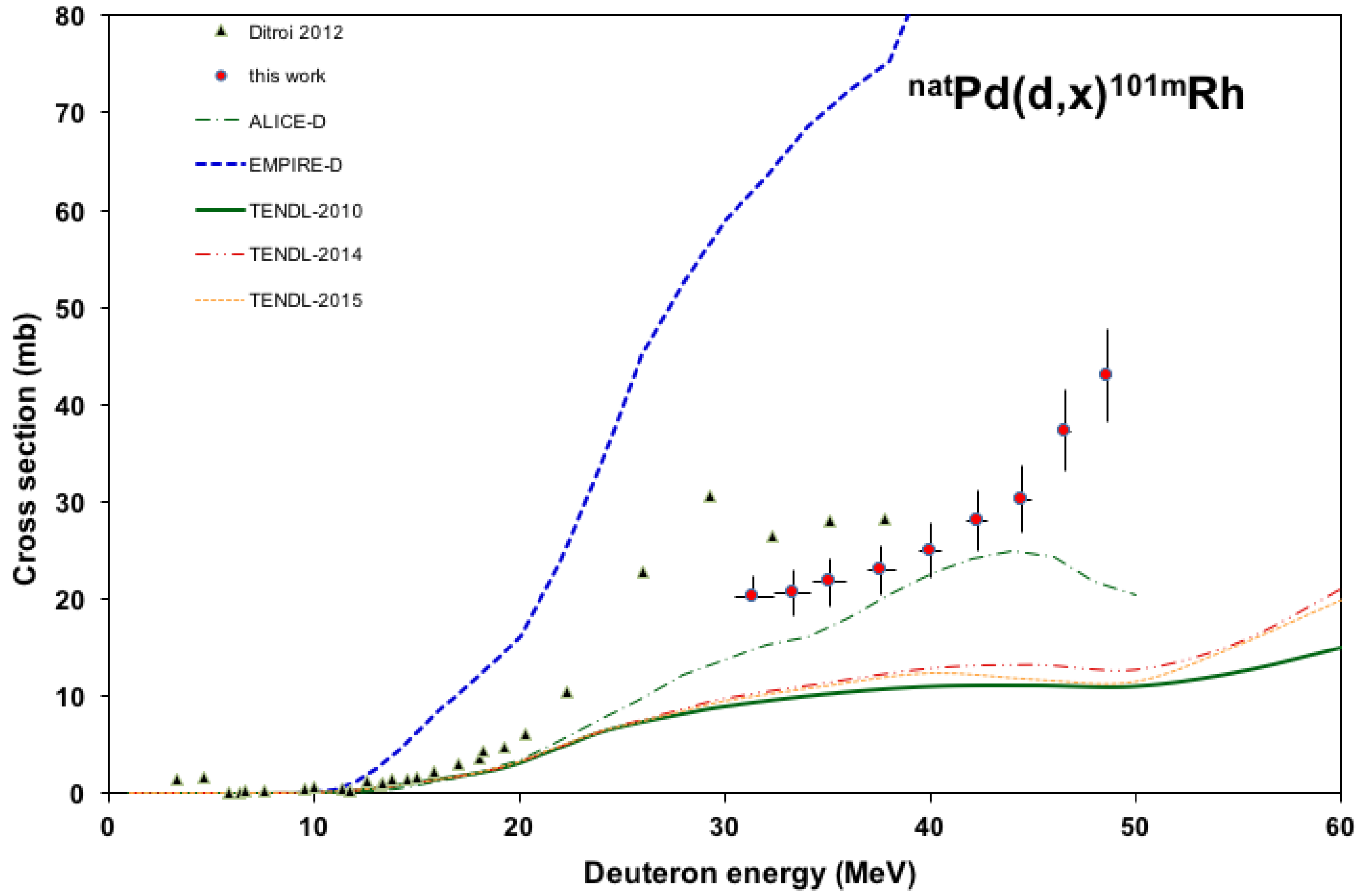

Fig. 14. Excitation functions of the $^{nat}Pd(d,x)^{101m}Rh$ reaction in comparison with literature values and theoretical results

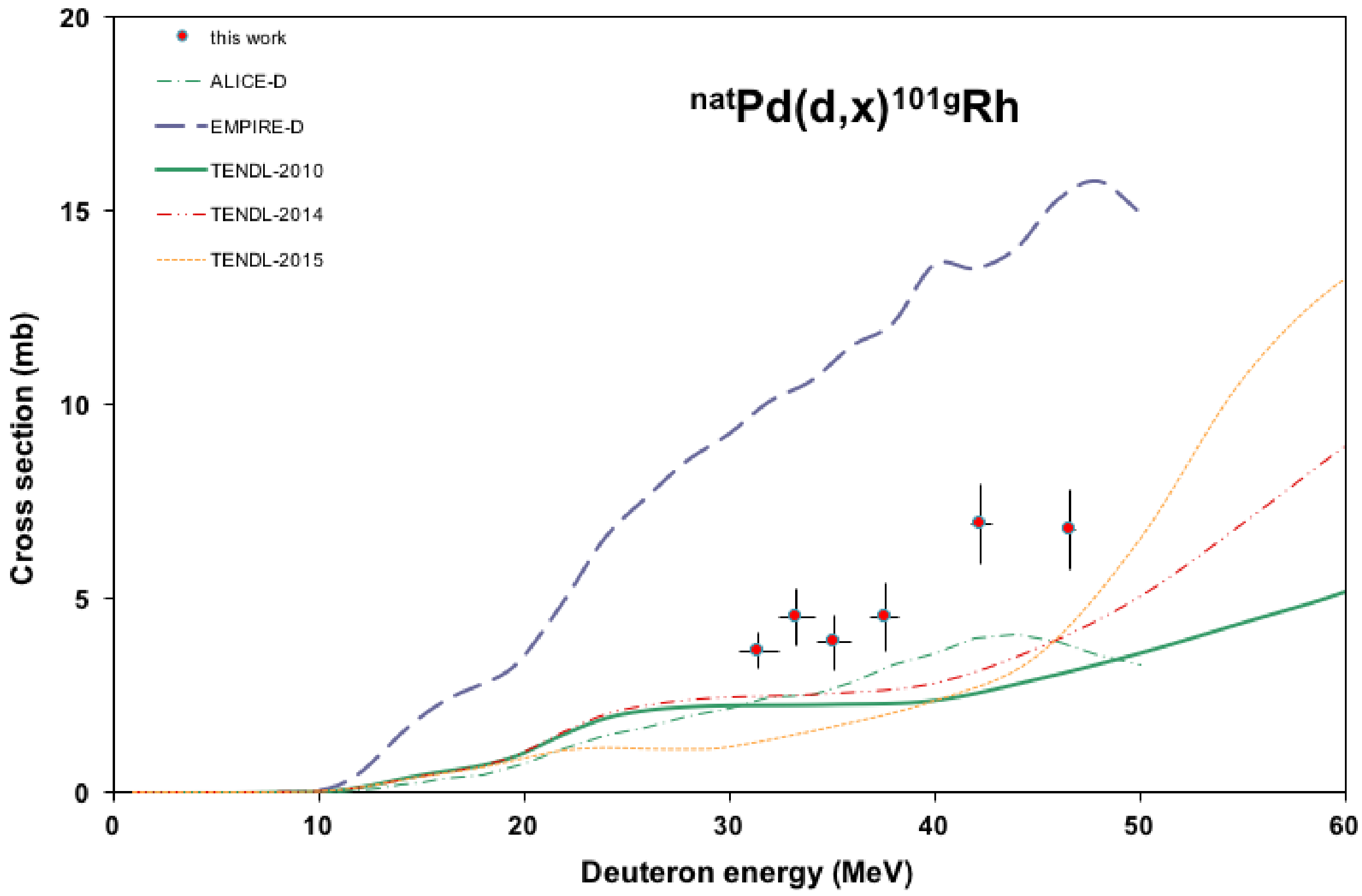

Fig. 15. Excitation functions of the $^{nat}Pd(d,x)^{101g}Rh$ reaction in comparison with literature values and theoretical results

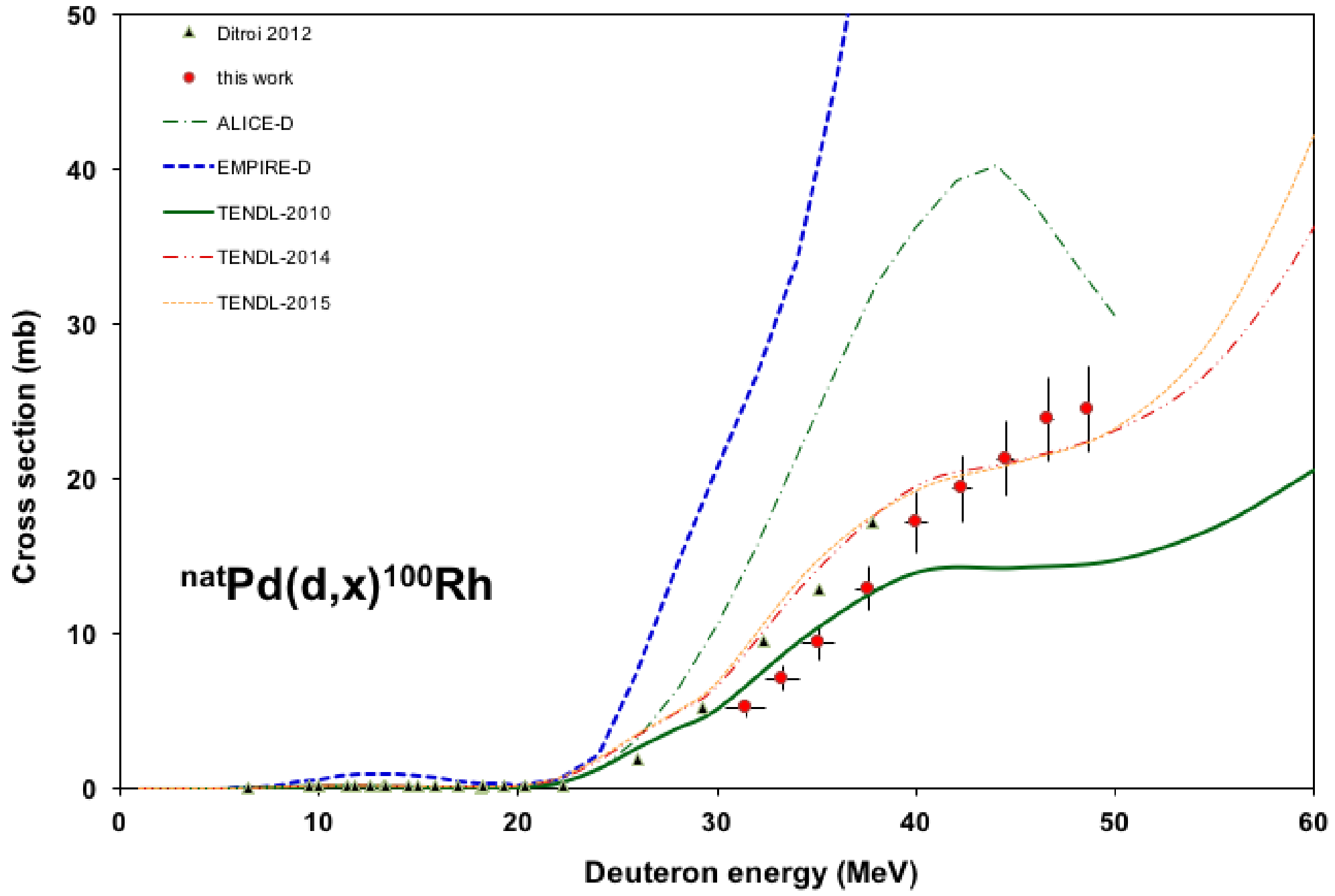

Fig. 16. Excitation functions of the $^{nat}Pd(d,x)^{100}Rh$ reaction in comparison with literature values and theoretical results

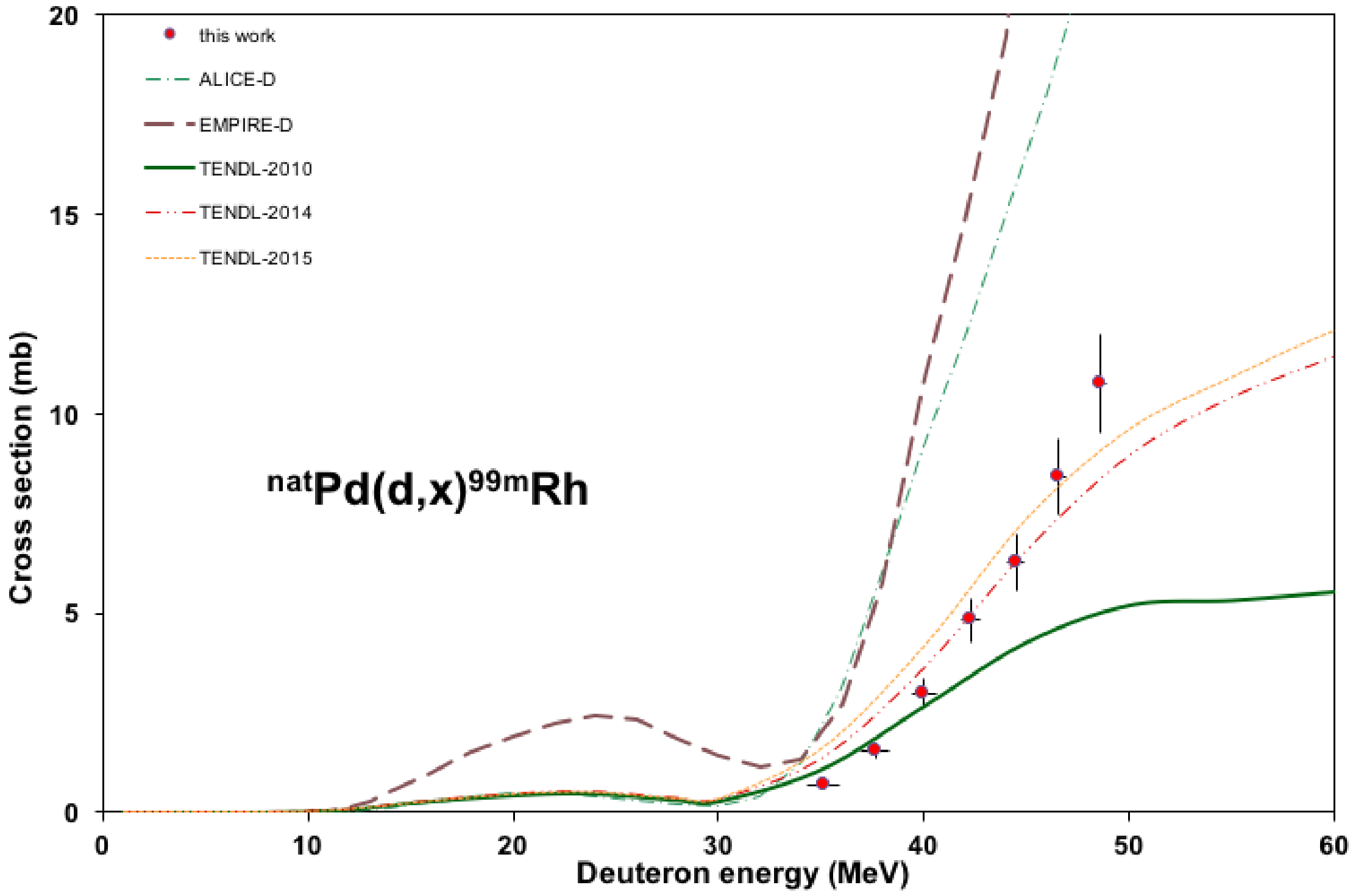

Fig. 17. Excitation functions of the $^{nat}Pd(d,x)^{99m}Rh$ reaction in comparison with literature values and theoretical results

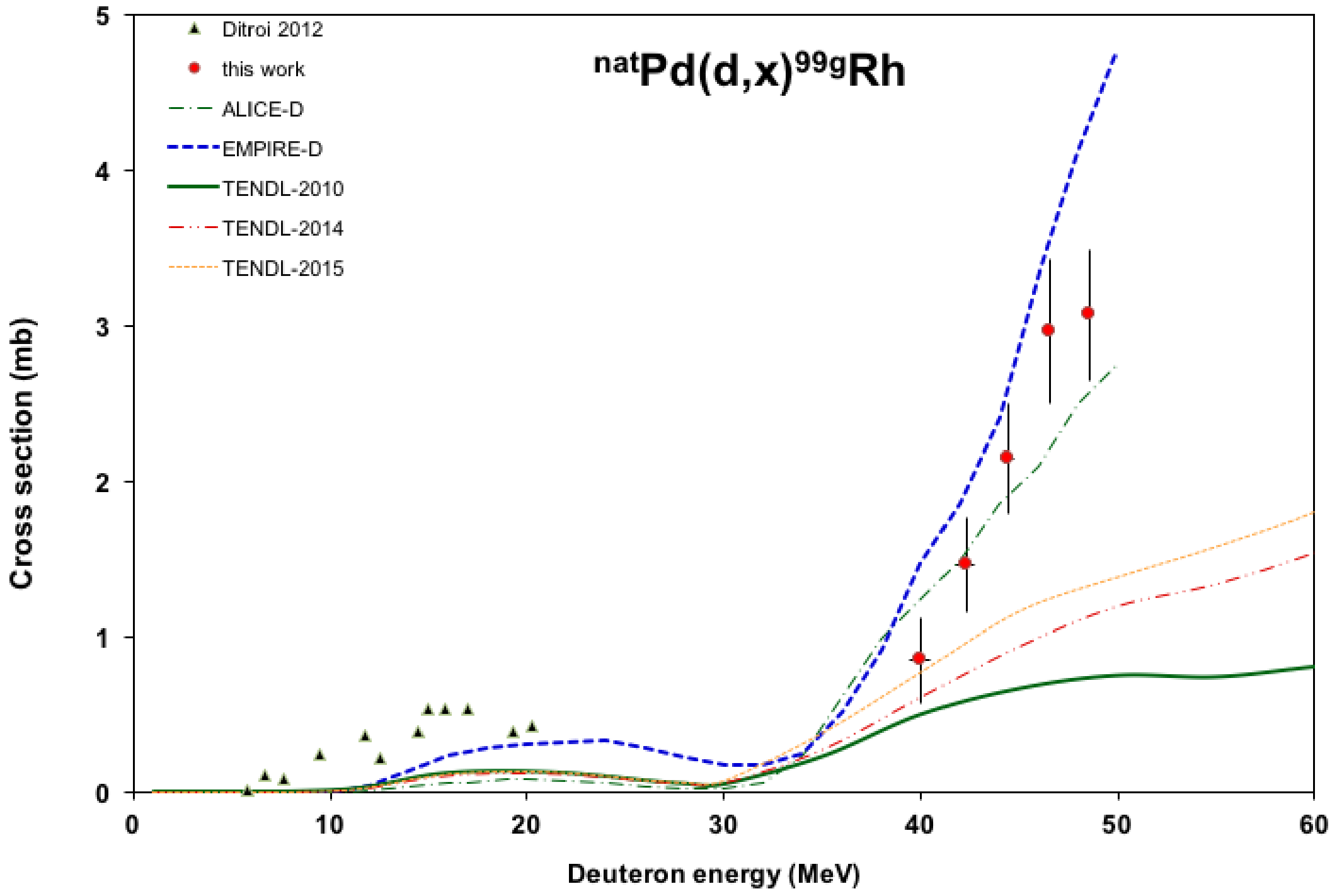

Fig. 18. Excitation functions of the $^{nat}Pd(d,x)^{99g}Rh$ reaction in comparison with literature values and theoretical results

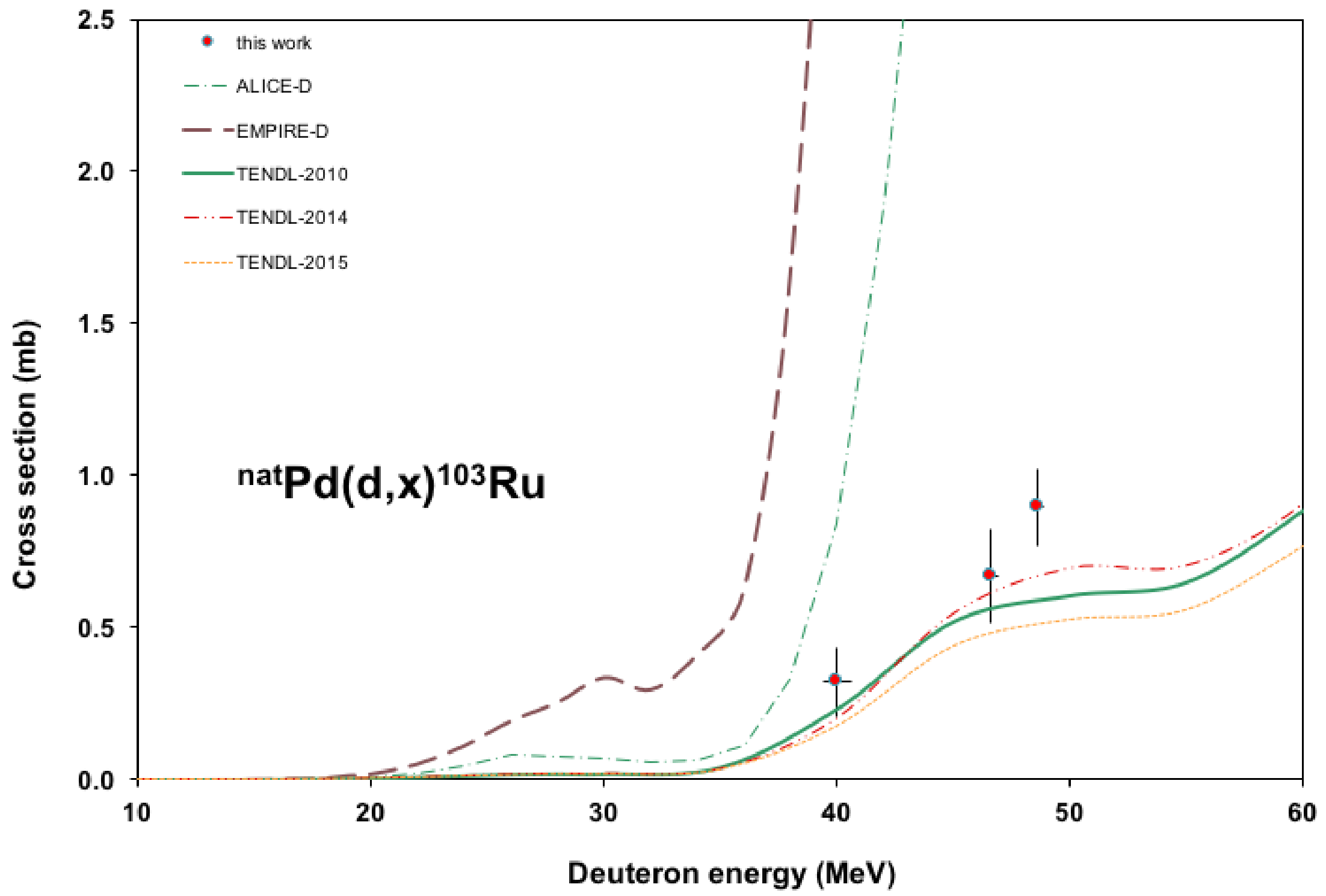

Fig. 19. Excitation functions of the $^{nat}Pd(d,x)^{103}Ru$ reaction in comparison with theoretical results

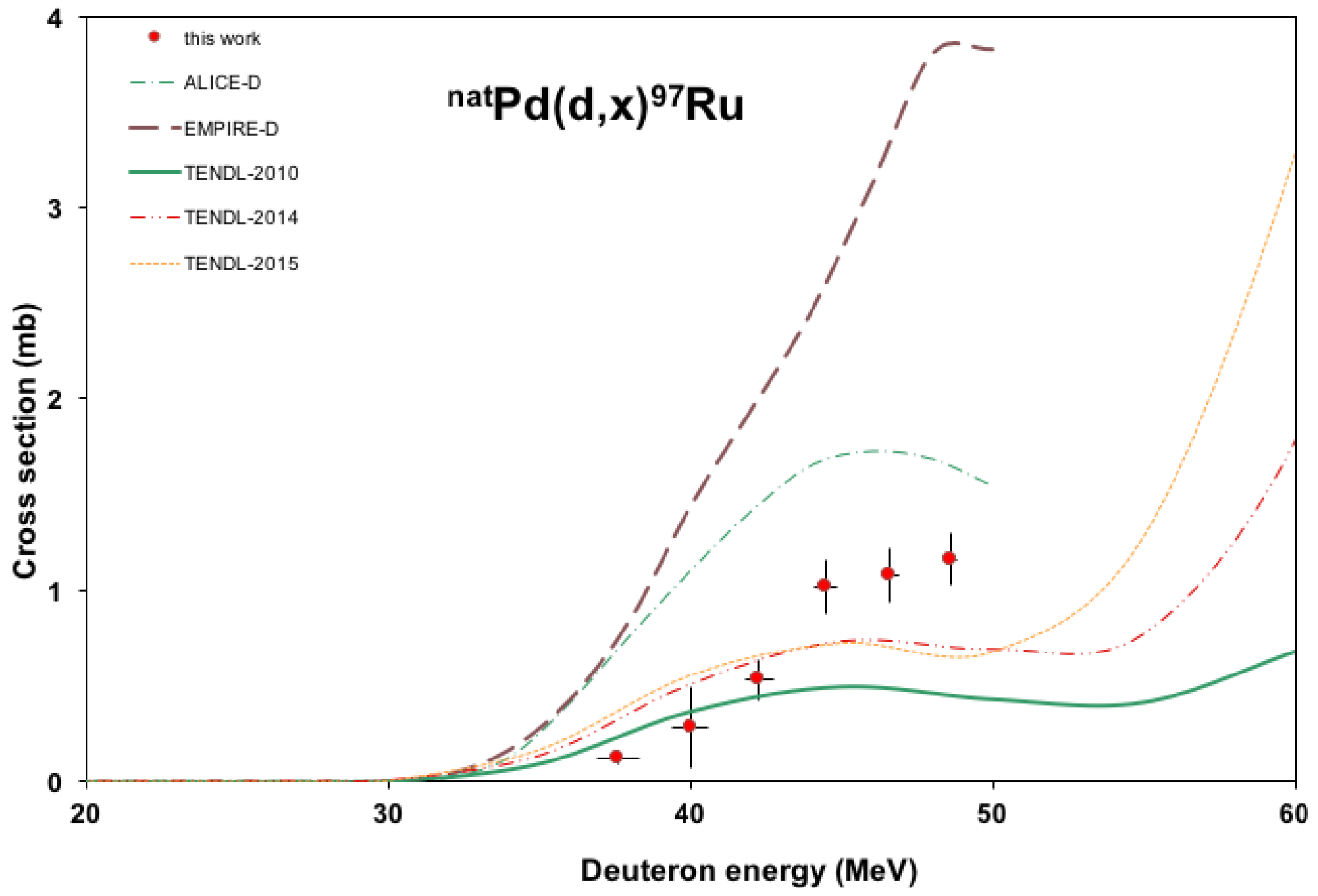

Fig. 20. Excitation functions of the $^{nat}Pd(d,x)^{97}Ru$ reaction in comparison with theoretical results

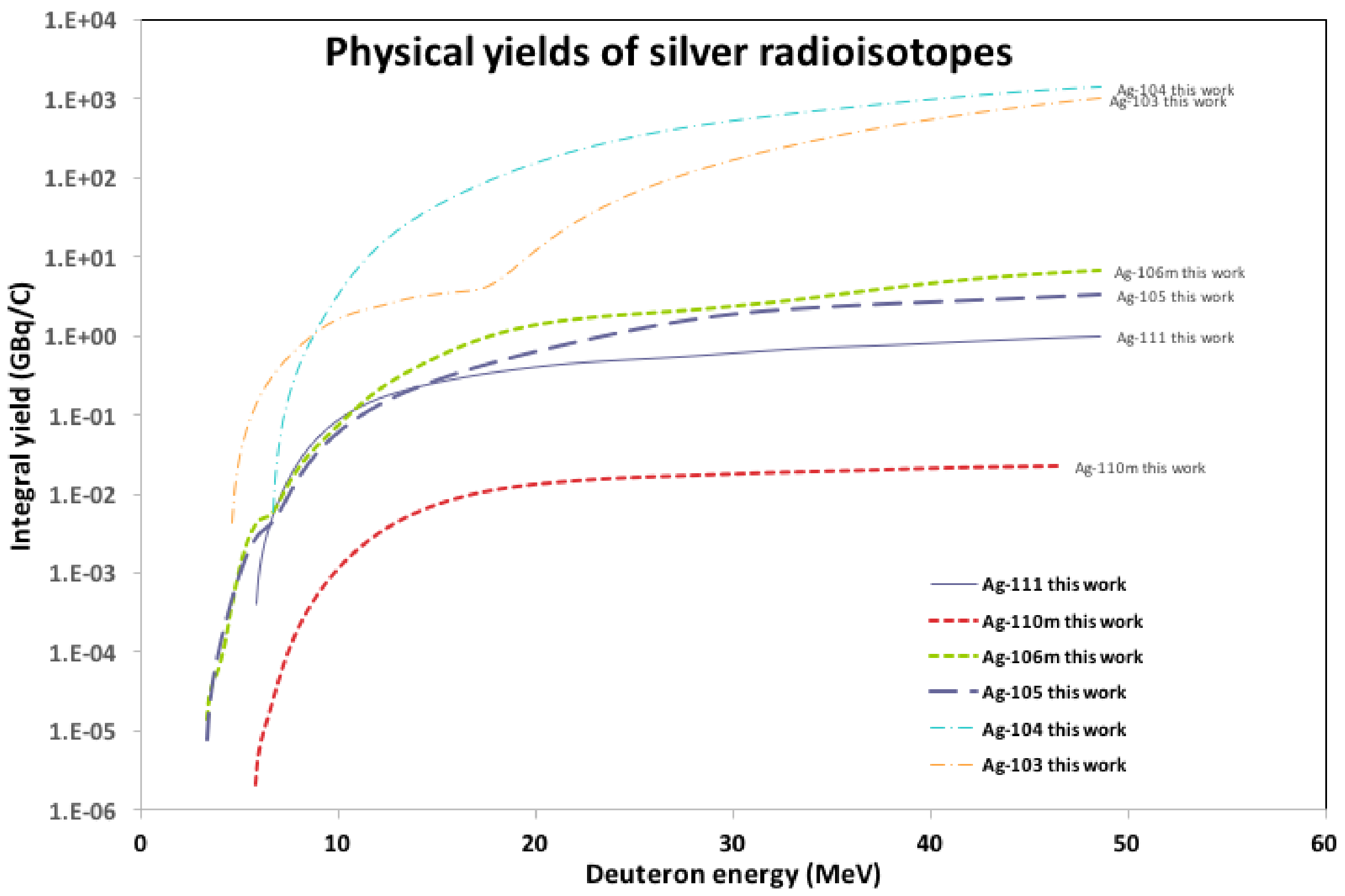

Fig. 21. Integral yields for the $^{nat}Pd(d,xn)^{111,110m,106m,105,104g,103}Ag$ reactions

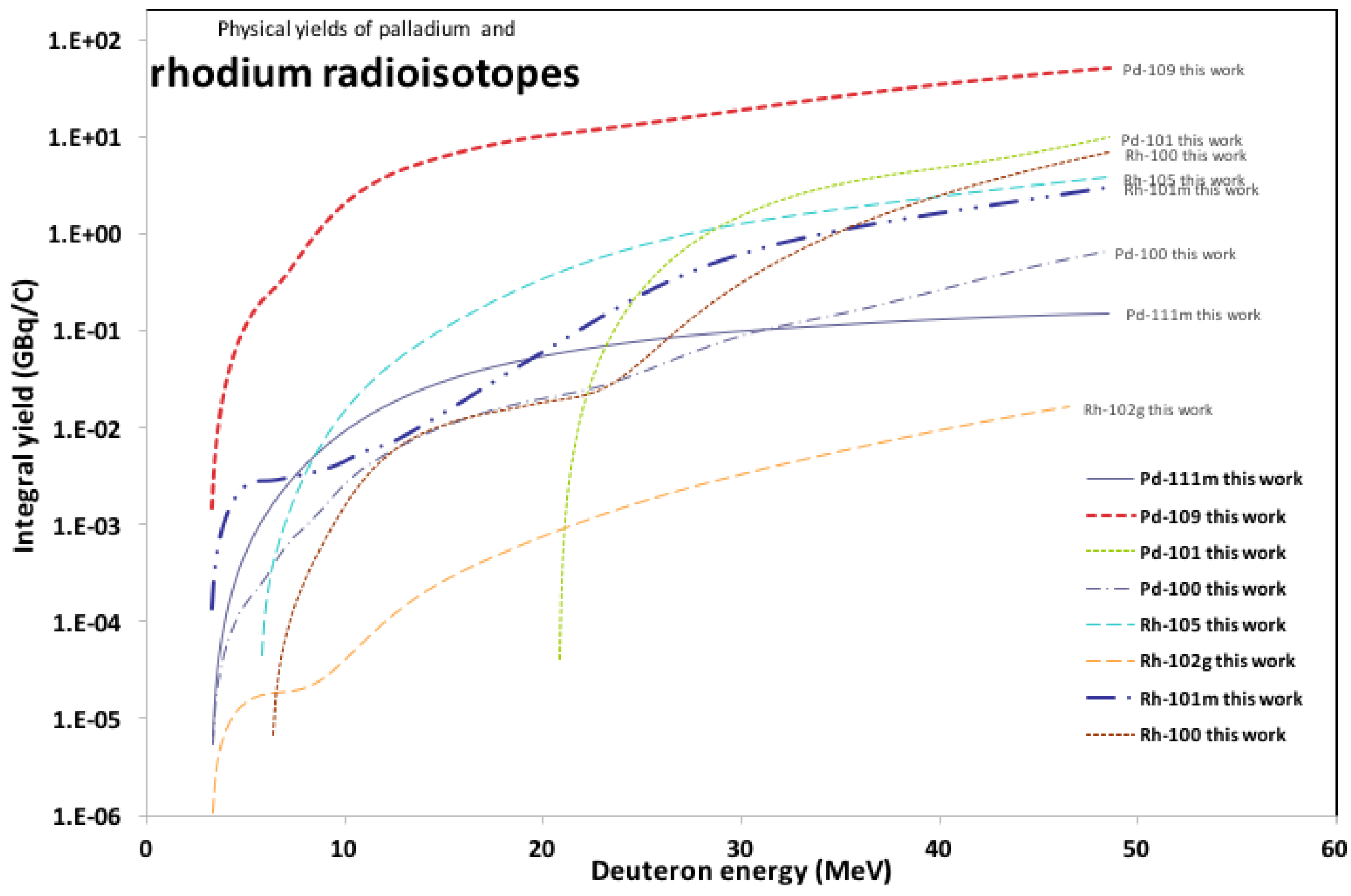

Fig. 22. Integral yields for the $^{nat}Pd(d,x)^{111m,109,101,100}Pd$, $^{nat}Pd(d,x)$, $^{105,102g,101m,100}Rh$ reactions

# References

Andersen, H.H., Ziegler, J.F., 1977. Hydrogen stopping powers and ranges in all elements. The stopping and ranges of ions in matter, Volume 3. Pergamon Press, New York.

Bonardi, M., 1987. The contribution to nuclear data for biomedical radioisotope production from the Milan cyclotron facility, in: Okamoto, K. (Ed.), Consultants Meeting on Data Requirements for Medical Radioisotope Production. IAEA, INDC(NDS)-195 (1988), Tokyo, Japan, p. 98.

Ditrói, F., Tárkányi, F., Ali, M.A., 2000. Investigation of deuteron induced nuclear reactions on niobium. Nucl. Instrum. Methods Phys. Res., Sect. B 161, 172-177.

Ditrói, F., Tárkányi, F., Takács, S., Mahunka, I., Csikai, J., Hermanne, A., Uddin, M.S., Hagiwara, M., Baba, M., Ido, T., Shubin, Y., Dityuk, A.I., 2007. Measurement of activation cross sections of the proton induced nuclear reactions on palladium. J. Radioanal. Nucl. Chem. 272, 231-235.

Ditrói, F., Tárkányi, F., Takács, S., Hermanne, A., Ignatyuk, A.V., Baba, M., 2012. Activation cross-sections of deuteron induced reactions on natural palladium. Nucl. Instrum. Methods Phys. Res., Sect. B 270, 61-74.

Dityuk, A.I., Konobeyev, A.Y., Lunev, V.P., Shubin, Y.N., 1998. New version of the advanced computer code ALICE-IPPE, INDC (CCP)-410. IAEA, Vienna.

Herman, M., Capote, R., Carlson, B.V., Oblozinsky, P., Sin, M., Trkov, A., Wienke, H., Zerkin, V., 2007. EMPIRE: Nuclear reaction model code system for data evaluation. Nucl. Data Sheets 108, 2655-2715.

Hermanne, A., Takács, S., Tárkányi, F., Bolbos, R., 2002. Cross section for the charged particle production of the therapeutic radionuclide Ag-111 and its PET imagins analogue Ag-104g, Annales Universitatis Turkuensis, Seria, Turku, Finland, p. 14.

Hermanne, A., Takács, S., Tárkányi, F., Bolbos, R., 2004a. Cross section measurements of proton and deuteron induced formation of Ag-103 in natural palladium. Radiochim. Acta 92, 215-218.

Hermanne, A., Takács, S., Tárkányi, F., Bolbos, R., 2004b. Experimental cross sections for charged particle production of the therapeutic radionuclide Ag-111 and its PET imaging analogue Ag-104m,Ag-g. Nucl. Instrum. Methods Phys. Res., Sect. B 217, 193-201.

Hermanne, A., Tárkányi, F., Takács, S., Shubin, Y.N., 2004c. Experimental determination of cross section of alpha-induced reactions on natPd, in: Haight, R.C., Talou, P., Kawano, T. (Eds.), International Conference on Nuclear Data for Science and Technology. AIP, Santa Fe, USA, pp. 961-964.

Hermanne, A., Tárkányi, F., Takács, S., Shubin, Y.N., 2005. Experimental determination of cross section of alpha-induced reactions on Pd-nat. Nucl. Instrum. Methods Phys. Res., Sect. B 229, 321-332.

Hermanne, A., Tárkányi, F., Takács, S., Ditrói, F., Baba, M., Ohtshuki, T., Spahn, I., Ignatyuk, A.V., 2009. Excitation functions for production of medically relevant radioisotopes in deuteron irradiations of Pr and Tm targets. Nucl. Instrum. Methods Phys. Res., Sect. B 267, 727-736.

Ignatyuk, A.V., 2010. 2nd RCM on FENDL-3. IAEA, Vienna, Austria,http://www-nds.iaea.org/fendl3/RCM2_slides.html.

Ignatyuk, A.V., 2011. Phenomenological systematics of the (d,p) cross sections, http://www-nds.iaea.org/fendl3/000pages/RCM3/slides//Ignatyuk_FENDL-3%20presentation.pdf, in: IAEA (Ed.), Vienna.

International-Bureau-of-Weights-and-Measures, 1993. Guide to the expression of uncertainty in measurement, 1st ed. International Organization for Standardization, Genève, Switzerland.

Koning, A.J., Rochman, D., 2010. TENDL-2010: "TALYS-based Evaluated Nuclear Data Library" ftp://ftp.nrg.eu/pub/www/talys/tendl2010/tendl2010.html.

Koning, A.J., Rochman, D., 2012. Modern Nuclear Data Evaluation with the TALYS Code System. Nucl. Data Sheets 113, 2841.

Koning, A.J., Rochman, D., van der Marck, S., Kopecky, J., Sublet, J.C., Pomp, S., Sjostrand, H., Forrest, R., Bauge, E., Henriksson, H., Cabellos, O., Goriely, S., Leppanen, J., Leeb, H., Plompen, A., Mills, R., 2014. TENDL-2014: TALYS-based evaluated nuclear data library, www.talys.eu/tendl2014.html.

Koning, A.J., Rochman, D., Kopecky, J., Sublet, J.C., Bauge, E., Hilaire, S., Romain, P., Morillon, B., Duarte, H., van der Marck, S., Pomp, S., Sjostrand, H., Forrest, R., Henriksson, H., Cabellos, O., S., G., Leppanen, J., Leeb, H., Plompen, A., Mills, R., 2015. TENDL-2015: TALYS-based evaluated nuclear data library,, https://tendl.web.psi.ch/tendl_2015/tendl2015.html.

NuDat, 2014. NuDat2 database (2.6). National Nuclear Data Center, Brookhaven National Laboratory. http://www.nndc.bnl.gov/nudat2/
Pritychenko, B., Sonzogni, A., 2003. Q-value calculator. NNDC, Brookhaven National Laboratory. http://www.nndc.bnl.gov/qcalc.

Székely, G., 1985. Fgm - a flexible gamma-spectrum analysis program for a small computer. Comput. Phys. Commun. 34, 313-324.

Tárkányi, F., Szelecsényi, F., Takács, S., 1991. Determination of effective bombarding energies and fluxes using improved stacked-foil technique. Acta Radiol., Suppl. 376, 72.

Tárkányi, F., Takács, S., Gul, K., Hermanne, A., Mustafa, M.G., Nortier, M., Oblozinsky, P., Qaim, S.M., Scholten, B., Shubin, Y.N., Youxiang, Z., 2001. Beam monitor reactions (Chapter 4). Charged particle cross-section database for medical radioisotope production: diagnostic radioisotopes and monitor reactions. , TECDOC 1211. IAEA, p. 49,http://www.nds.or.at/medical.

Tárkányi, F., Hermanne, A., Takács, S., Hilgers, K., Kovalev, S.F., Ignatyuk, A.V., Qaim, S.M., 2007. Study of the $^{192}$Os(d,2n) reaction for, production of the therapeutic radionuclide $^{192}$Ir in no-carrier added form. Appl. Radiat. Isot. 65, 1215-1220.

Tárkányi, F., Hermanne, A., Király, B., Takács, S., Ditrói, F., Csikai, J., Fenyvesi, A., Uddin, M.S., Hagiwara, M., Baba, M., Ido, T., Shubin, Y.N., Ignatyuk, A.V., 2009. New cross-sections for production of Pd-103; review of charged particle production routes. Appl. Radiat. Isot. 67, 1574-1581.

Tárkányi, F., Ditrói, F., Takács, S., Csikai, J., Hermanne, A., Uddin, M.S., Baba, M., 2016. Activation cross sections of proton induced nuclear reactions on palladium up to 80 MeV. Appl. Radiat. Isot. 114, 128-144.